# An Experimental Estimation Method of Diffusion Coefficients in Ternary and Multicomponent Systems from a Single Diffusion Couple Profile

Suman Sadhu[1], Anindita Chakraborty[2], Surendra Makineni[1], Saswata Bhattacharyya[3], Aloke Paul[1]*

[1]Department of Materials Engineering, Indian Institute of Science, Bengaluru, India
[2]Research and Development, Tata Steel, Jamshedpur, India
[3]Materials Science and Metallurgical Engineering, Indian Institute of Technology, Hyderabad, India
*Corresponding author: Email: aloke@iisc.ac.in

## Abstract

Until recently, it was textbook knowledge that the diffusion coefficients could not be estimated in a multi-component system following the widely practised diffusion couple method. The recently proposed constrained diffusion couple methods need two intersecting diffusion paths in a multi-component space with very well-controlled compositions of the diffusion couple end members, which may be tricky depending on the complications of diffusion paths in certain systems. In this study, we have proposed a method for estimating all types of diffusion coefficients directly at the Kirkendall marker plane from a single diffusion couple. The estimation method and design strategy for producing diffusion couples are discussed in concentrated ternary Ni-Co-Fe, Fe-rich quaternary Fe-Ni-Co-Cr and Ni-rich Ni-Co-Fe-Cr-Al quinary alloys. As demonstrated further, one can even estimate the impurity diffusion coefficients utilizing the composition profiles at the ends of the diffusion couples, which has been rarely practised in multicomponent systems until now. We have further demonstrated the importance of estimating the tracer and intrinsic diffusion coefficients in a concentrated or multi-principal element alloy in which interdiffusion coefficients can be vague and misleading for understanding the elements' diffusional interactions and relative mobilities. We have also shown the importance of considering the vacancy wind effect in concentrated alloys. The method and design strategy of producing diffusion couples demonstrated in this study can be suitable for generating a mobility database in Ni-, Co-based (superalloys) and Fe-based (steel) multicomponent systems with relative ease, which was considered impossible until recently.

## 1. Introduction

Estimating diffusion coefficients following the widely practised diffusion couple method in a multicomponent system was considered one of the most important but unsolved challenges until recently [1–3]. The diffusion couple technique is the oldest and most used method for estimating diffusion coefficients. On the other hand, the radiotracer method was considered the only option in multicomponent systems [4–7] because of unsolved complications related to the diffusion couple method. However, safety issues and the non-availability of affordable radioisotopes of many important elements used in various applications make it impossible to practice the radiotracer method as the only option [1]. Therefore, solving the issues with the diffusion couple method in multicomponent systems is essential for generating the mobility database. Many have dealt with this differently over the last several decades without successfully estimating meaningful diffusion coefficients.

Estimating all types of diffusion coefficients (tracer, intrinsic and interdiffusion coefficients) in binary systems following the diffusion couple method is relatively straightforward [3,8,9]. Most importantly, these can be estimated from a single diffusion couple. On the other hand, two diffusion paths must be intersected to estimate the interdiffusion coefficients and then estimate the tracer and intrinsic diffusion coefficients in the ternary system [3,10]. However, estimating the intrinsic diffusion coefficients directly at the Kirkendall marker plane is almost impossible [1,3]. Complications increase rapidly with a further increase in the number of elements in the system. Estimating diffusion coefficients following the diffusion couple method in a n-component system with $n > 3$ was considered impossible since it is almost impossible to intersect $(n-1)$ (or even two) conventional diffusion paths (in which all the elements develop the diffusion profiles) in a multicomponent space [1,3].

Recently, Morral introduced the concept of the body diagonal diffusion couple method. He proposed a specific design strategy of diffusion couples in a composition range of (nearly) constant diffusion coefficients such that diffusion paths may not intersect but pass closely to estimate the interdiffusion coefficients [11]. However, identifying a small composition range with constant interdiffusion coefficients such that (n-1) diffusion paths pass very closely to estimate the interdiffusion coefficients directly can be a painstaking task since this range can differ for different alloy systems. Morral proposed this mainly for estimating interdiffusion coefficients in multi-principal element alloy systems, which was practised experimentally by Verma et al. [12]. However, as discussed in this article, the interdiffusion coefficients, especially in multi-principal element alloy, can be vague or misleading unless the tracer and intrinsic diffusion coefficients are calculated to understand the relative mobilities and actual diffusional interactions between the elements. Moreover, the range of errors because of considering (n-1) diffusion paths passing with different distances is unknown. Therefore, Das and Paul demonstrated that the same method could be extended to estimate the tracer diffusion coefficients directly from only two closely passed diffusion paths and then calculate the intrinsic and interdiffusion coefficients irrespective of the number of elements in a system [13]. This can still be practised in solid solutions where diffusion coefficients do not vary significantly in many systems over a relatively small composition range. However, this method may be challenging to practice in various intermetallic compounds, for example, B2 MX-based (M= Ni, Co, Fe, Au, etc. and X= Al, Zn, Ga, etc.) multicomponent systems of practical importance in which ternary [14,15] and multicomponent characteristics of diffusion paths can be very complicated compared to the solid solutions. Even a reasonable composition range of constant diffusion coefficient is difficult to find since diffusion coefficients vary significantly with composition because of the very high content of constitutional defects in these types of compounds [15–17].

On the other hand, the constrained diffusion couple methods established by the group of Paul, such as the pseudo-binary [18], pseudo-ternary [19] diffusion couple methods and the method of intersecting dissimilar diffusion paths [20-22] are shown to be helpful in both multicomponent solid solution [20–22] as well as intermetallic compounds [15,17]. The concepts of intersecting pseudo-ternary and dissimilar diffusion paths broke the notion that the diffusion paths cannot be intersected in multicomponent space, which hindered the diffusion community for estimation of the diffusion in multicomponent systems during the last several decades [19-20]. The diffusion rates of only two and three elements could be estimated using pseudo-binary and pseudo-ternary diffusion couple methods, respectively. However, the diffusion coefficients of all the elements can be estimated by intersecting only two diffusion profiles following the concept of intersecting dissimilar diffusion paths [20–23], which has an immense benefit. For example, over the last several decades, the diffusion community has been engaged with a mindset to estimate the $(n-1)^2$ interdiffusion coefficients first, which needs (n-1) diffusion paths to be intersected. This was proposed even by Kirkaldy and Lane for estimating the tracer diffusion coefficients in a ternary system after estimating the interdiffusion coefficients [10]. This created all the difficulties. By moving away from this frame of thinking, Paul's group proposed the estimation of the tracer diffusion coefficients directly from the interdiffusion flux [1]. This needs only two diffusion paths to intersect for estimating $n$ tracer diffusion coefficients by correlating $2(n-1)$ interdiffusion fluxes directly with tracer diffusion coefficients, and then solving the equations following the least square method [13,20-22]. This brings an immense benefit in higher-order systems, leading to the ease of designing experiments.

Further significant advantages can be achieved if the tracer, intrinsic and interdiffusion diffusion coefficients can also be estimated from a single diffusion profile purely experimentally instead of two intersecting or closely passed diffusion profiles even in a multicomponent system. We cannot estimate the $n(n-1)$ intrinsic diffusion coefficients directly at the Kirkendall marker plane since we need $(n-1)$ diffusion paths to intersect, and additionally, the marker plane must be present at the composition of intersection in all the diffusion couples, which is impossible to achieve. By changing the mindset again, we demonstrate in this article, for the first time, that we can estimate the tracer diffusion coefficients of all the elements directly at the Kirkendall marker plane utilizing the known thermodynamic parameters in ternary and multicomponent systems, most importantly, from a single diffusion profile irrespective of the number of elements in a system. Following this, we can also calculate the intrinsic and interdiffusion coefficients. We have first demonstrated this in the Ni-Co-Fe ternary system, which has gained importance as one of the important medium entropy alloy systems and is also important for various high-temperature material systems. The data estimated at the Kirkendall marker plane from a single diffusion couple is compared with those estimated at the intersection of two diffusion paths following the Kirkaldy Lane method [10] by passing two diffusion paths near the Kirkendall marker plane. This helps establish the reliability of estimated data from a single diffusion couple by producing reliable diffusion profiles. We have then extended this to the higher-order systems. We have first demonstrated this method in a quaternary Fe-rich FeNiCoCr system such that we can compare the data estimated at the Kirkendall marker plane with the estimated impurity diffusion coefficients of Ni, Co and Cr in pure Fe by the Hall method [24,25] and tracer diffusion coefficient of Fe in pure Fe estimated by radiotracer method [26]. We have then extended our analysis to the quinary Ni-rich NiCoFeCrAl alloy system related to the NiCoFeCrAl multi-principal element alloy, one of the most studied high-entropy alloys. Following, we have also discussed questioning the usefulness of the interdiffusion coefficients, which, in general, are attempted to estimate. We have shown that the interdiffusion coefficients in a material system with relatively low alloying elements can be useful since the interdiffusion coefficients are almost equal to the intrinsic diffusion coefficients. However, the interdiffusion coefficients can be vague and misleading to understand the diffusional interactions, especially in concentrated and multi-principal (high entropy alloys). In such cases, we need to estimate tracer and intrinsic diffusion coefficients, which reflect rightly on the relative mobilities of the elements in the absence of thermodynamic driving forces and diffusional interactions between the elements in the presence of thermodynamic driving forces. Therefore, the method proposed in this study can be very useful for generating the mobility database in multicomponent systems of practical importance following the diffusion couple method with relative ease, which was considered very difficult or almost impossible until recently. The need for just one diffusion couple, irrespective of the number of elements in the system, has immense benefits for the ease of generation of mobility databases in various multicomponent systems.

## 2. Experimental method

The end member alloys for producing diffusion couples were melted with pure Ni, Co, Fe, Cr and Al (99.95 - 99.99 wt.%) under an argon atmosphere in a vacuum arc melting unit. All the alloy buttons were remelted five times and annealed in a vacuum furnace (~ $10^{-4}$ Pa) at 1200 °C for 50 h to achieve compositional homogeneity. The average composition of the alloy buttons was measured by spot analysis at various random positions on the metallographically polished buttons using WDS (wavelength dispersive spectroscopy) in EPMA (electron probe micro analyser). The compositional variation is found to be within 0.15 at. % from the average composition. The target and actual compositions of different end-member alloys are listed in Table 1. Slices of ~1.5 mm were cut from the buttons using EDM (electro-discharge machine). After metallographic preparation, $Y_2O_3$ particles (0.5-1 $\mu m$) dispersed in acetone were spread on one of the bonding interfaces. $Y_2O_3$ particles were evenly distributed after evaporating the acetone, which helped to detect the Kirkendall marker plane in the diffusion couple [3]. The diffusion couples were assembled in a special fixture and subjected to diffusion annealing at 1200°C for 50 h in a vacuum, followed by water quenching. The diffusion couples were then cross-sectioned by a slow-speed diamond saw and prepared metallographically for microstructure analysis and WDS line profile measurement in the EPMA.

Table 1. Diffusion couple end member compositions: target and actual compositions

| Diffusion Couple | End Members | Target Composition | Actual Composition |
|---|---|---|---|
| DC1 Fe-$Ni_{50}Co_{50}$ | EM1 | Fe | Fe |
| | EM2 | $Ni_{50}Co_{50}$ | $Ni_{50}Co_{50}$ |
| DC2 Ni-$Fe_{50}Co_{50}$ | EM1 | Ni | Ni |
| | EM2 | $Fe_{50}Co_{50}$ | $Fe_{49.8}Co_{50.2}$ |
| DC3 Co-($Fe_{50}Ni_{50}$ | EM1 | Co | Co |
| | EM2 | $Fe_{50}Ni_{50}$ | $Fe_{50.8}Ni_{49.2}$ |
| DC4 Fe-$Ni_{62.5}Co_{37.5}$ | EM1 | Fe | Fe |
| | EM2 | $Ni_{62.5}Co_{37.5}$ | $Ni_{61.9}Co_{38.1}$ |
| DC5 Co-$Fe_{37.5}Ni_{62.5}$ | EM1 | Co | Co |
| | EM2 | $Ni_{62.5}Fe_{37.5}$ | $Ni_{61.9}Fe_{38.1}$ |
| DC6 Fe-$Fe_{67.5}Ni_{7.5}Co_{7.5}Cr_{7.5}$ | EM1 | Fe | Fe |
| | EM2 | $Fe_{77.5}Ni_{7.5}Co_{7.5}Cr_{7.5}$ | $Fe_{77.2}Ni_{7.4}Co_{7.7}Cr_{7.7}$ |
| DC7 Ni-$Ni_{23.75}Co_{23.75}Fe_{23.75}Cr_{23.75}Al_5$ | EM1 | Ni | Ni |
| | EM2 | $Ni_{23.75}Co_{23.75}Fe_{23.75}Cr_{23.75}Al_5$ | $Ni_{23.4}Co_{23.8}Fe_{24.1}Cr_{24.1}Al_{4.6}$ |

## 3. Results and Discussion

### *3.1 Estimation of diffusion coefficients from a single diffusion profile in ternary and multicomponent systems*

Several diffusion studies are conducted in the binary Ni-Co, Co-Fe, and Ni-Fe systems because of the importance of this material system for high-temperature applications [27–37]. Several studies during the last several decades have also been conducted in the Ni-Co-Fe ternary systems to estimate the interdiffusion coefficients or understand the characteristics of the diffusion profiles [38–40]. This system has gained renewed interest recently, considering it as one of the medium entropy alloy systems. However, for the first time in this study, we aim to estimate all types of diffusion coefficients purely experimentally from one diffusion couple and then compare the data estimated at the intersecting diffusion paths of two diffusion couples.

The interdiffusion flux of element *i* ($\tilde{J}_i$) is related to the interdiffusion coefficients ($\widetilde{D}_{ij}^{n}$) following [3,41]

$$V_m\tilde{J}_i = -\sum_{j=1}^{n-1} \widetilde{D}_{ij}^{n} \frac{\partial N_j}{\partial x} \quad (1)$$

In a ternary system, $(n-1)^2 = 4$ interdiffusion coefficients considering, let's say Fe as the dependent variable, are related to two independent interdiffusion fluxes by [[3,41]]

$$V_m\tilde{J}_{Ni} = -\widetilde{D}_{NiNi}^{Fe} \frac{\partial N_{Ni}}{\partial x} - \widetilde{D}_{NiCo}^{Fe} \frac{\partial N_{Co}}{\partial x} \quad (2a)$$

$$V_m\tilde{J}_{Co} = -\widetilde{D}_{CoNi}^{Fe} \frac{\partial N_{Ni}}{\partial x} - \widetilde{D}_{CoCo}^{Fe} \frac{\partial N_{Co}}{\partial x} \quad (2b)$$

where $\widetilde{D}_{NiNi}^{Fe}$ and $\widetilde{D}_{CoCo}^{Fe}$ are the main interdiffusion coefficients, $\widetilde{D}_{NiCo}^{Fe}$ and $\widetilde{D}_{CoNi}^{Fe}$ are the cross interdiffusion coefficients. $\frac{\partial N_j}{\partial x}$ is the composition gradient when composition $N_j$ is expressed in atomic or mole fraction in a metallic system. $V_m$ is the molar volume considered constant in a ternary and multicomponent system. The interdiffusion flux of the element of interest can be calculated directly from the composition profile over the whole interdiffusion zone by [3,42,43]

$$V_m\tilde{J}_i = -\frac{N_i^+ - N_i^-}{2t}\left[(1 - Y_i^*)\int_{x^{-\infty}}^{x^*} Y_i dx + Y_i^* \int_{x^*}^{x^{+\infty}} (1 - Y_i) dx\right] \quad (3)$$

where $N_i^-$ and $N_i^+$ are the left- and right-hand side of the unaffected part of the diffusion couple. $Y_i = \frac{N_i - N_i^-}{N_i^+ - N_i^-}$ is the composition normalized variable and $t$ is annealing time at the temperature of interest. Therefore, one can directly estimate the interdiffusion coefficients at the intersecting composition of two diffusion paths by writing four equations from two diffusion couples.

Similarly, the intrinsic flux of element *i* ( $J_i$) are related to intrinsic diffusion coefficients ($D_{ij}^{n}$) by [3, 41]

$$V_m J_i = -\sum_{j=1}^{n-1} D_{ij}^{n} \frac{\partial N_j}{\partial x} \quad (4)$$

Therefore $n(n-1) = 6$ intrinsic diffusion coefficients considering Fe as the dependent variable can be expressed as

$$V_m J_{Ni} = -D^{Fe}_{NiNi}\frac{\partial N_{Ni}}{\partial x} - D^{Fe}_{NiCo}\frac{\partial N_{Co}}{\partial x} \quad (5a)$$

$$V_m J_{Co} = -D^{Fe}_{CoNi}\frac{\partial N_{Ni}}{\partial x} - D^{Fe}_{CoCo}\frac{\partial N_{Co}}{\partial x} \quad (5b)$$

$$V_m J_{Cr} = -D^{Fe}_{CrNi}\frac{\partial N_{Ni}}{\partial x} - D^{Fe}_{CrCo}\frac{\partial N_{Co}}{\partial x} \quad (5c)$$

Unlike interdiffusion fluxes, the intrinsic fluxes can be calculated directly only at the Kirkendall marker plane ($x_K$) by [3]

$$V_m J_i = -\frac{1}{2t}\left[N_i^+ \int_{x^{-\infty}}^{x_K} Y_i dx - N_i^- \int_{x_K}^{x^{+\infty}} (1 - Y_i)\, dx\right] \quad (6)$$

Therefore, again we need two diffusion paths to intersect for writing six equations for estimation of six intrinsic diffusion coefficients. However, the Kirkendall marker planes are not necessarily located at the composition of intersection in both the diffusion couples. This can be understood from three diffusion couples produced in the Ni-Co-Fe system, as shown in Fig. 1. These are produced by coupling Fe/NiCo (DC 1), Ni/CoFe (DC 2), and Co/NiFe (DC 3). Following, these are plotted in Gibbs triangle to highlight the composition of intersections between the diffusion paths and the locations of the Kirkendall marker planes, as shown in Fig. 2. It can be seen that the Kirkendall marker planes ($K_1, K_2$, and $K_3$) are located at very different compositions compared to the intersecting compositions of the diffusion paths. Therefore, we can estimate the interdiffusion coefficients directly at the intersecting compositions but cannot estimate the intrinsic diffusion coefficients directly at the Kirkendall marker planes. To solve this problem, we can express the intrinsic fluxes of elements directly with the tracer diffusion coefficient. The intrinsic and tracer diffusion coefficients are related by [44]

$$D^n_{ij} = \frac{N_i}{N_j} D^*_i \emptyset^n_{ij}\left(1 + W^n_{ij}\right) \text{ where } W^n_{ij} = \frac{2}{S_o \emptyset^n_{ij}} \frac{\sum_{i=1}^{n} N_i D^*_i \emptyset^n_{ij}}{\sum_{i=1}^{n} N_i D^*_i} \quad (7)$$

$W^n_{ij}$ is the vacancy wind effect as proposed by Manning considering the cross terms of the Onsager parameters [45,46]. $\emptyset^n_{ij} = \frac{\partial ln a_i}{\partial ln N_1} - \frac{N_1}{N_n}\frac{\partial ln a_i}{\partial ln N_n}$ is the thermodynamic parameter expressed with activity (*a*) of elements [44]. Therefore, by replacing Eq. 7 in 4, the intrinsic fluxes can be directly related to the tracer diffusion coefficients by

$$V_m J_i = -\sum_{j=1}^{n-1}\left[\frac{N_i}{N_j} D^*_i \emptyset^n_{ij}\left(1 + W^n_{ij}\right)\right]\frac{\partial N_j}{\partial x} \quad (8)$$

One can understand that if we neglect the vacancy wind effect in Eq. 7 (*i.e.* $W^n_{ij} = 0$), intrinsic flux of an element, *i* is related to the tracer diffusion coefficient of the same element and the thermodynamic factors. If the vacancy wind effect cannot be neglected, then the intrinsic flux of an element is related to the tracer diffusion coefficients of all the elements. Since we noted a significant contribution of the vacancy wind effect on certain intrinsic diffusion coefficients in this system at the composition estimated (which is explained in Section 3.3), we have estimated the tracer diffusion coefficients considering the vacancy wind effect. The thermodynamic factors at these compositions are calculated from the activity data available in data bank TCFE9 and TCNI9 of ThermoCalc [47–50], which are given in the supplementary file. The estimated tracer diffusion coefficients by calculating the intrinsic fluxes of elements at the Kirkendall marker positions in the Ni-Co-Fe system are listed in Table 2.

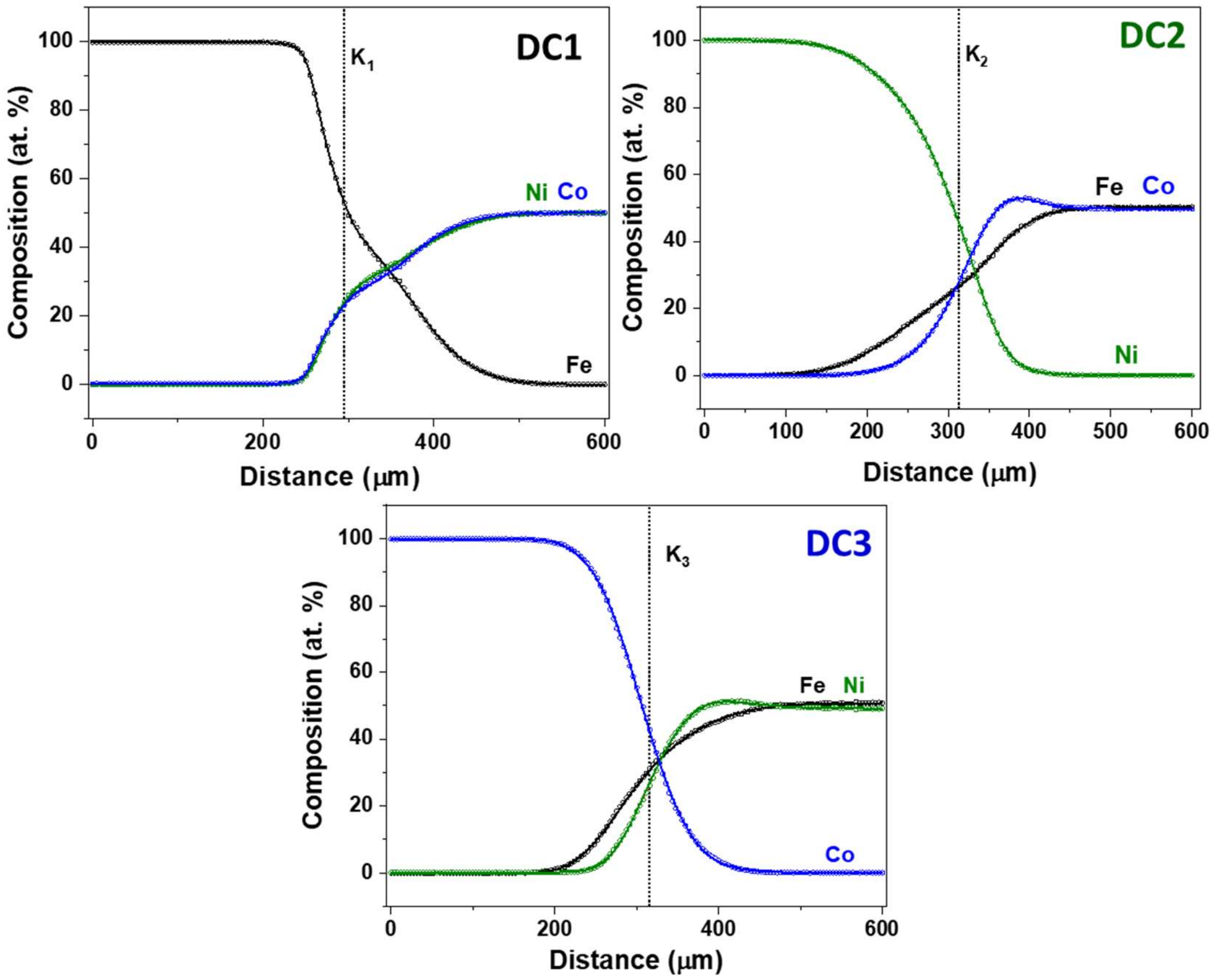

Fig.1. The diffusion profiles produced at 1200°C after annealing for 50 h in three diffusion couples designated as DC1, DC2, DC3 and their corresponding Kirkendall marker plane are marked as $K_1$, $K_2$ and $K_3$ respectively.

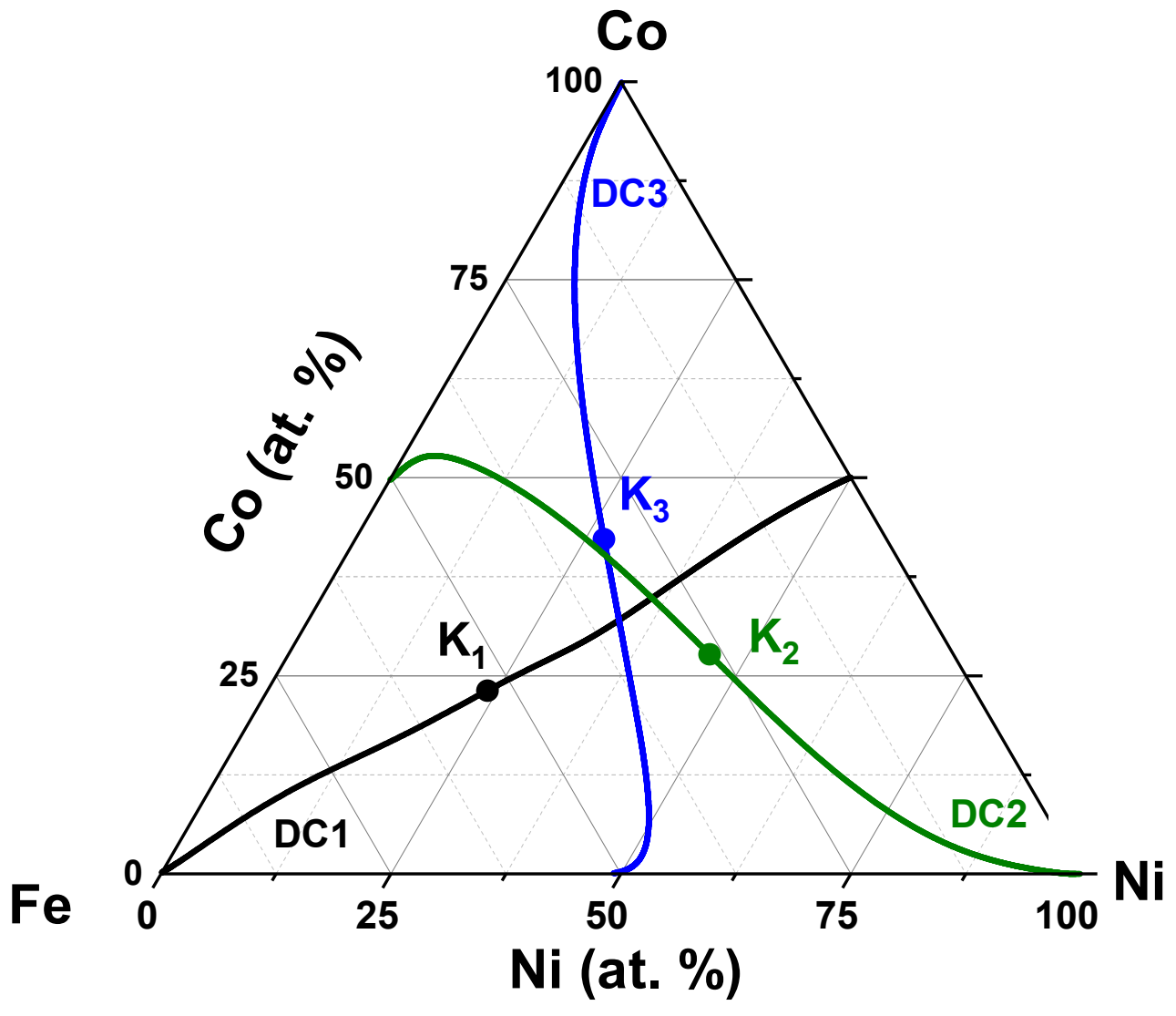

Fig.2. Diffusion paths produced at 1200°C in the diffusion couples DC1, DC2 and DC3 C are plotted on Gibbs triangle.

Table 2. Tracer diffusion coefficients estimated at the Kirkendall Marker planes at 1200°C

| Marker Plane Location Composition (at. %) | Tracer Diffusion Coefficients ($\times 10^{-15}$ m $^2$/s) | | |
|---|---|---|---|
| | $D^*_{Fe}$ | $D^*_{\mathrm{Ni}}$ | $D^*_{\mathrm{Co}}$ |
| DC 1 ($K_1$) (Fe = 53.0 Ni = 23.9 Co = 23.1) | 12.1±2.4 | 4.9±1 | 4.7±0.9 |
| DC 2 ($K_2$) (Fe = 26.5 Ni = 45.8 Co = 27.7) | 20.7±4.5 | 6.9±1.4 | 6.5±1.3 |
| DC 3 ($K_3$) (Fe = 30.7 Ni = 27.1 Co = 42.2) | 10.3±2.1 | 4.9±1.1 | 5.4±0.9 |

One important point should be noted here about the diffusion couple design strategy we followed in this ternary system with complete solid solution. The diffusion couples are produced such that the composition of all the elements is zero in either left- or right-hand side of the diffusion couple (*i.e.* $N_i^- = 0$ or $N_i^+ = 0$). Therefore, the intrinsic fluxes of elements at the Kirkendall marker plane can be calculated from $V_m J_i = -\frac{1}{2t}\left[N_i^+ \int_{x^{-\infty}}^{x_K} Y_i dx\right]$ or $V_m J_i = -\frac{1}{2t}\left[N_i^- \int_{x_K}^{x^{+\infty}} (1 - Y_i)\, dx\right]$. This reduces the error in calculation significantly. In an incremental diffusion couple, when $N_i^- \neq 0$ and $N_i^+ \neq 0$, the calculated tracer diffusion coefficients values may induce a higher error and may be solved with illogical negative value of tracer diffusion coefficient because of experimental error and nature of the Eq. 6 (negative sign inside the bracket). These problems are well known in binary systems. An error in locating the Kirkendall marker plane or accurate composition of the end members or even the inferior quality of the diffusion couples may lead to a big error in calculation of the tracer and intrinsic diffusion coefficients [3,51,52]. In multicomponent systems, most of the phases have limited solubility range, and depending on the complexity of the phase diagrams. Therefore, one may need to produce the incremental diffusion couple, in which the quality of the diffusion couple produced and composition profile measured have utmost importance. Additionally, a certain design strategy of such incremental diffusion couples is important, which we have demonstrated in quaternary and quinary alloys after the analysis in this Ni-Co-Fe system.

To compare with the data estimated at the Kirkendall marker plane, we followed the Kirkaldy-Lane method [10,53] by intersecting the diffusion paths at compositions close to the Kirkendall marker planes. As shown in Fig. 3, two more diffusion couples DC 4 and DC 5 are produced such that these two intersects with DC 2 at compositions close to the composition of Kirkendall marker plane, $K_2$ of this diffusion couple. Kirkaldy and Lane proposed to estimate the interdiffusion coefficients at the intersecting composition of the diffusion paths utilizing Eq. 1-3. Two equations can be written from each couple following Eq. 2 and a total of four equations from two couples can be solved to estimate the interdiffusion coefficients directly. Further, the interdiffusion coefficients are related to the intrinsic diffusion coefficients at this composition (proposed by extending the relations established in the binary system based on the Kirkendall marker experiment [54-56]) by [3, 41]

$$\widetilde{D}_{ij}^{n} = D_{ij}^{n} - N_i\left(\sum_{k=1}^{n} D_{kj}^{n}\right) \tag{9}$$

Replacing Eq. 7 in Eq. 9, we can express the interdiffusion coefficients directly with the tracer diffusion coefficients following,

$$\widetilde{D}_{ij}^{n} = \frac{N_i}{N_j} D_i^* \emptyset_{ij}^{n}\left(1 + W_{ij}^{n}\right) - N_i\left[\sum_{k=1}^{n} \frac{N_k}{N_j} D_i^* \emptyset_{kj}^{n}\left(1 + W_{kj}^{n}\right)\right] \tag{10}$$

Therefore, we can calculate three tracer diffusion coefficients following the least square method from four already estimated interdiffusion coefficients in this ternary system. It should be noted here that Kirkaldy-Lane [10] proposed this correlation neglecting the vacancy wind effect and chemical potential of elements. The group of van Loo [53] proposed this correlation with thermodynamic factors (which are materials constants similar to the diffusion coefficients) but neglecting the vacancy wind factor, although they commented on the probable importance of this factor. We noticed a significant influence of the vacancy wind effect especially in multi-principal element alloy on certain cross-intrinsic diffusion coefficients (as discussed in section 3.3) and decided to calculate the data in Ni-Co-Cr system in which the data are estimated at significant concentration of all the elements.

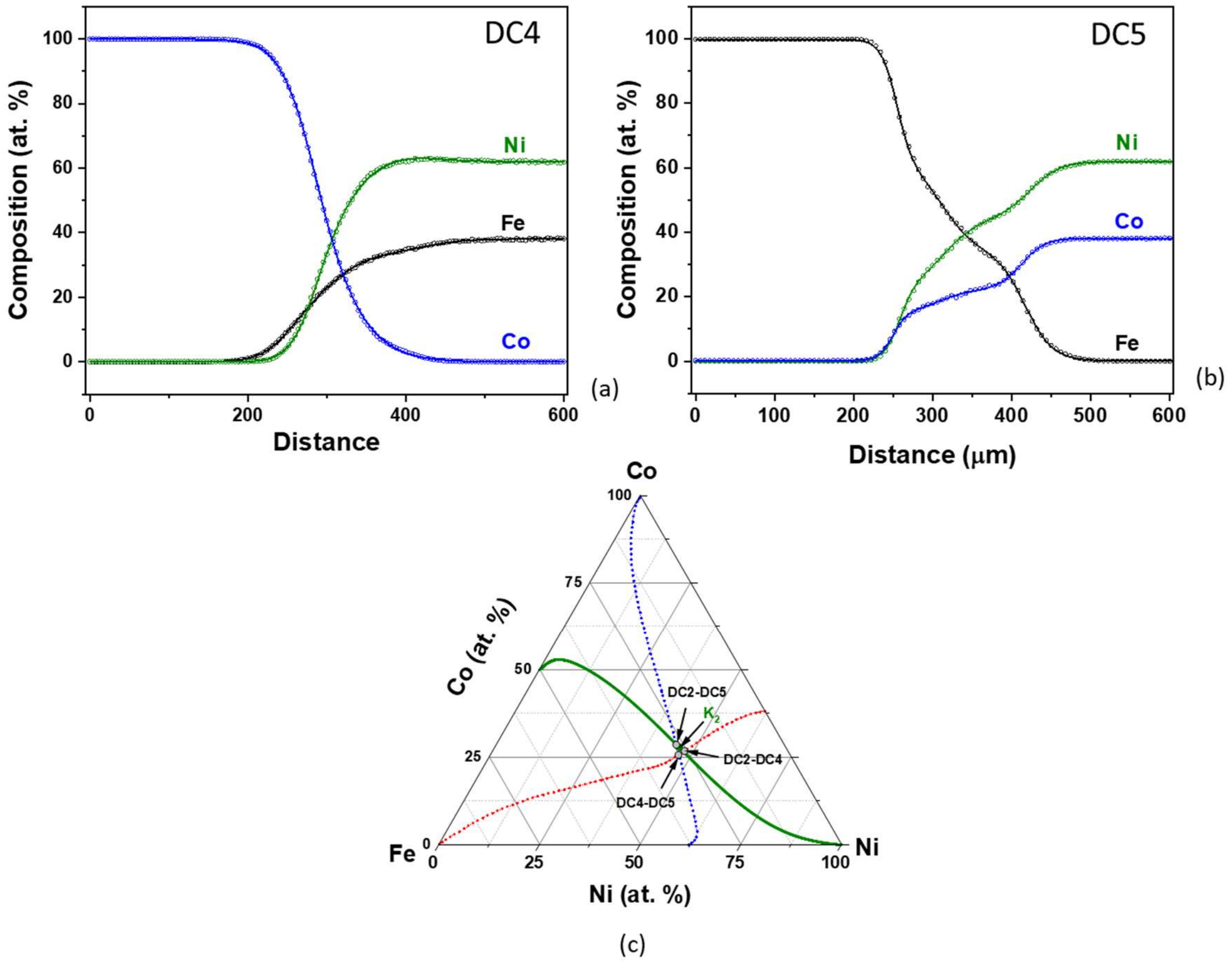

Fig. 3 Two diffusion couples (a) DC4 and (b) DC5 are produced at 1200°C after annealing for 50 h. (c) Three intersecting compositions of the diffusion couples (DC2-DC5, DC2-DC4 and DC4-DC5) at very close composition of the Kirkendall marker plane of DC2 ($K_2$).

The estimated diffusion coefficients at the Kirkendall marker plane ($K_2$) directly following the method proposed in this study and at the intersecting compositions following the Kirkaldy-Lane method are compared in Table 3a. The data estimated following two different methods estimated at four close compositions are found to be very similar. Following, we can calculate the intrinsic diffusion coefficients utilizing Eq. 7 and then calculate the interdiffusion coefficients utilizing Eq. 9. The directly estimated interdiffusion coefficients at the intersecting compositions of DC2, DC4 and DC5 and the interdiffusion coefficients calculated from the estimated tracer diffusion coefficients at the Kirkendall marker plane $K_2$ are listed in Table 3b for comparison to find comparable values. Therefore, we can indeed estimate all the diffusion coefficients directly at the Kirkendall marker from a single diffusion couple for generation of reliable mobility database.

Table 3 Comparison of (a) tracer diffusion coefficients and (b) interdiffusion coefficients estimated at Kirkendall marker plane K2 (DC2) and the intersecting compositions of DC2-DC4, DC2-DC5, DC4-DC5.

| Tracer Diffusion Coefficients ($\times 10^{-15}$ m $^2$/s) | $K_2$<br>Fe - 26.1<br>Ni - 46.0<br>Co -27.9 | DC 2 – DC 4<br>Fe - 26.1<br>Ni - 47.5<br>Co - 26.4 | DC 2 – DC 5<br>Fe = 26.8<br>Ni = 44.7<br>Co = 28.5 | DC 4 – DC 5<br>Fe = 27.6<br>Ni = 46.8<br>Co = 25.6 |
|---|---|---|---|---|
| $D^*_{Fe}$ | 20.7±4.5 | 16.0±4.1 | 16.0±3.2 | 18.8±3.7 |
| $D^*_{Ni}$ | 6.9±1.4 | 7.7±1.5 | 8.2±1.6 | 6.8±1.3 |
| $D^*_{Co}$ | 6.5±1.3 | 6.6±1.4 | 5.5±1.2 | 7.0±1.4 |

(a)

| | Direct Estimation of Interdiffusion Coefficients ($\times 10^{-15}$ m $^2$/s) | | | Interdiffusion coefficients at the Kirkendall Marker Plane ($\times 10^{-15}$ m $^2$/s) |
|---|---|---|---|---|
| | DC2-DC4 | DC2-DC5 | DC4-DC5 | K2 (DC2) |
| $\tilde{D}^{Fe}_{NiNi}$ | 15.4±3.1 | 15.1±3 | 17.2±3.2 | 17.8 |
| $\tilde{D}^{Fe}_{CoNi}$ | 4.9±1 | 4.8±0.9 | 6.6±1,2 | 6.9 |
| $\tilde{D}^{Fe}_{NiCo}$ | 8.0±1.5 | 8.2±1.6 | 9.6±1.9 | 11.6 |
| $\tilde{D}^{Fe}_{CoCo}$ | 10.4±2 | 10.1±1.9 | 11.6±2.1 | 13.0 |

(b)

A similar comparison of the tracer diffusion coefficients estimated following the method proposed in this study at $K_3$ in DC3 and the data estimated at the intersecting composition of DC2 and DC3 following the Kirkaldy Lane method is given in the supplementary file (S1). As an additional set of data, we can estimate the diffusion coefficients at three intersecting compositions of DC1-DC2, DC1-DC3 and DC2-DC3, which are given in the supplementary file (S2 and S3).

Now let us demonstrate the estimation of diffusion coefficients from a single diffusion couple in a quaternary system producing the incremental diffusion couple, which is very important since a phase may have limited solubility of alloying elements in various multicomponent system unlike the Ni-Co-Fe system at the temperature interest described already. In such a situation, we have no option but produce an incremental diffusion couple in which the diffusion couple cannot be produced with $N_i^- = 0$ or $N_i^+ = 0$ for the major (parent) element although this can be fulfilled for the minor elements for estimation of data with relatively smaller range of error if the diffusion couple is designed in a certain way. For example, the diffusion couple in the Fe-rich Fe-Ni-Co-Cr quaternary system is produced by coupling Fe with Fe7.5Ni7.5Co7.5Cr alloy (in atomic percentage). The composition profile measured in this diffusion couple is shown in Fig. 4. Therefore, we have $N_i^- = 0$ for Ni, Co and Cr. However, $N_i^- \neq 0$ or $N_i^+ \neq 0$ for Fe, which may induce a higher range of error in calculation which is discussed based on the estimated tracer diffusion coefficients. It is common practice to consider the solvent as the dependent variable in an alloy with relatively low concentration of alloying elements since the diffusion rate of these alloying elements controls the interdiffusion process. The correlation between intrinsic fluxes and intrinsic diffusion coefficients in this system considering substrate Fe as the dependent variable can be expressed as

$$V_m J_{Ni} = -D_{NiNi}^{Fe} \frac{\partial N_{Ni}}{\partial x} - D_{NiCo}^{Fe} \frac{\partial N_{Co}}{\partial x} - D_{NiCr}^{Fe} \frac{\partial N_{Cr}}{\partial x} \quad (11a)$$

$$V_m J_{Co} = -D_{CoNi}^{Fe} \frac{\partial N_{Ni}}{\partial x} - D_{CoCo}^{Fe} \frac{\partial N_{Co}}{\partial x} - D_{CoCr}^{Fe} \frac{\partial N_{Cr}}{\partial x} \quad (11b)$$

$$V_m J_{Cr} = -D_{CrNi}^{Fe} \frac{\partial N_{Ni}}{\partial x} - D_{CrCo}^{Fe} \frac{\partial N_{Co}}{\partial x} - D_{CrCr}^{Fe} \frac{\partial N_{Cr}}{\partial x} \quad (11c)$$

$$V_m J_{Fe} = -D_{FeNi}^{Fe} \frac{\partial N_{Ni}}{\partial x} - D_{FeCo}^{Fe} \frac{\partial N_{Co}}{\partial x} - D_{FeCr}^{Fe} \frac{\partial N_{Cr}}{\partial x} \quad (11d)$$

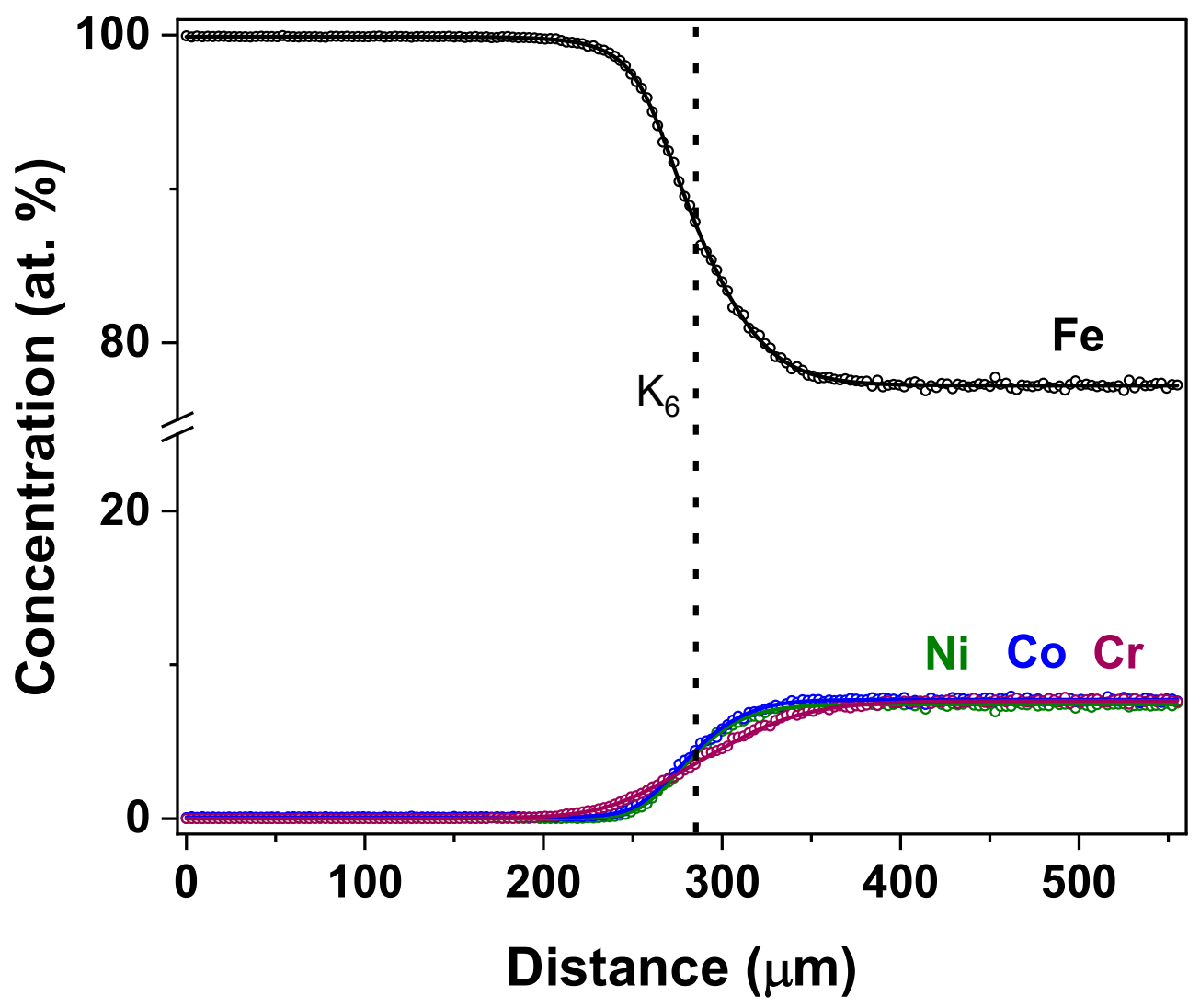

Fig. 4. Quaternary diffusion profile, produce at 1200°C after annealing for 50 h. The Kirkendall marker plane location is indicated by dotted line at $K_6$.

We can now express the intrinsic flux equations directly with tracer diffusion coefficients by replacing Eq. 7 in Eq. 11. We noted the negligible contribution of the vacancy wind effect in this composition of the alloy system (this is discussed in Section 3.3). Therefore, by considering $W_{ij}^{n} = 0$ in Eq. 9, Eq. 11 can be expressed as

$$V_m J_{Ni} = -D_{Ni}^{*} \emptyset_{NiNi}^{Fe} \frac{\partial N_{Ni}}{\partial x} - \frac{N_{Ni}}{N_{Co}} D_{Ni}^{*} \emptyset_{NiCo}^{Fe} \frac{\partial N_{Co}}{\partial x} - \frac{N_{Ni}}{N_{Cr}} D_{Ni}^{*} \emptyset_{NiCr}^{Fe} \frac{\partial N_{Cr}}{\partial x} \quad (12a)$$

$$V_m J_{Co} = -\frac{N_{Co}}{N_{Ni}} D_{Co}^{*} \emptyset_{CoNi}^{Fe} \frac{\partial N_{Ni}}{\partial x} - D_{Co}^{*} \emptyset_{CoCo}^{Fe} \frac{\partial N_{Co}}{\partial x} - \frac{N_{Co}}{N_{Cr}} D_{Co}^{*} \emptyset_{CoCr}^{Fe} \frac{\partial N_{Cr}}{\partial x} \quad (12b)$$

$$V_m J_{Cr} = -\frac{N_{Cr}}{N_{Ni}} D_{Cr}^{*} \emptyset_{CrNi}^{Fe} \frac{\partial N_{Ni}}{\partial x} - \frac{N_{Cr}}{N_{Co}} D_{Cr}^{*} \emptyset_{CrCo}^{Fe} \frac{\partial N_{Co}}{\partial x} - D_{Cr}^{*} \emptyset_{CrCr}^{Fe} \frac{\partial N_{Cr}}{\partial x} \quad (12c)$$

$$V_m J_{Fe} = -\frac{N_{Fe}}{N_{Ni}} D_{Fe}^{*} \emptyset_{FeNi}^{Fe} \frac{\partial N_{Ni}}{\partial x} - \frac{N_{Fe}}{N_{Co}} D_{Fe}^{*} \emptyset_{FeCo}^{Fe} \frac{\partial N_{Co}}{\partial x} - \frac{N_{Fe}}{N_{Cr}} D_{Fe}^{*} \emptyset_{FeCr}^{Fe} \frac{\partial N_{Cr}}{\partial x} \quad (12d)$$

Table 4 (a) Tracer diffusion coefficients estimated in diffusion couple D6 (Fe-rich, FeNiCoCr quaternary alloy) at 1200°C (b) Error in calculation if the location of the Kirkendall marker plane (K6) is detected at $\pm 1$ or $\pm 2\ \mu m$ from the position detected in this study.

| Marker Plane Location Composition (at. %) | Tracer Diffusion Coefficients ($\times 10^{-15}\ m^2/s$) | | | |
|---|---|---|---|---|
| | $D_{Fe}^{*}$ | $D_{Ni}^{*}$ | $D_{Co}^{*}$ | $D_{Cr}^{*}$ |
| DC 6 ($K_6$)<br>(Fe = 87.8 Ni = 4.2 Co = 4.5 Cr = 3.5 ) | 2.6±0.4 | 1.9±0.3 | 2.0±0.3 | 5.1±1.1 |

(a)

| Distance from K6 | Tracer Diffusion Coefficient ($\times 10^{-15}\ m^2/s$) | | | | % of error | | | |
|---|---|---|---|---|---|---|---|---|
| | $D_{Fe}^{*}$ | $D_{Ni}^{*}$ | $D_{Co}^{*}$ | $D_{Cr}^{*}$ | $D_{Fe}^{*}$ | $D_{Ni}^{*}$ | $D_{Co}^{*}$ | $D_{Cr}^{*}$ |
| -2 µm | 3.6 | 1.7 | 1.8 | 4.9 | 39.6 | -11.8 | -11.4 | -4.2 |
| -1 µm | 3.0 | 1.8 | 1.9 | 5.0 | 17.7 | -5.4 | -5.0 | -1.6 |
| K6 position | 2.6 | 1.9 | 2.0 | 5.1 | - | - | - | - |
| + 1 µm | 2.1 | 2.0 | 2.1 | 5.2 | -18.9 | 4.8 | 5.3 | 1.3 |
| + 2 µm | 1.3 | 2.1 | 2.2 | 5.2 | -48.4 | 11.2 | 11.0 | 1.4 |

(b)

Therefore, when the vacancy wind effect is negligible, the intrinsic flux of an element is related to the tracer diffusion coefficient of the same element only. This makes analysis much simpler. Further, since $N_i^- = 0$ for Ni, Co and Cr, the intrinsic flux of these elements can be calculated from $V_m J_i = -\frac{1}{2t}\left[N_i^+ \int_{x^{-\infty}}^{x_K} Y_i dx\right]$ (in reference to Eq. 6). However, the intrinsic flux of Fe needs to be calculated from Eq. 6. Therefore, a small error in locating the Kirkendall marker plane does not induce much error in calculated data of Ni, Co, Cr. However, a higher error can be induced for the calculation of Fe tracer diffusion coefficient. Of course, overall reliability of estimated data depends on the good quality of the diffusion couple produced, which otherwise will estimate non-reliable diffusion coefficients. To extend thus discussion, the tracer diffusion coefficients estimated are listed in Table 4a. The location of the marker plane can be accurately detected within $< 1\ \mu m$ of error with careful analysis. However, let us calculate the data for the situation we have uncertainty of $\pm 1\ \mu m$ and $\pm 2\ \mu m$ in locating the Kirkendall marker plane. It can be seen in Fig. 4b that Ni and Co induce similar range of difference although within the typical range of error of diffusion studies since these two elements have similar diffusion profile lengths. Cr induced even smaller range of difference since it has a higher diffusion length. Overall, these elements induced relatively smaller difference since the diffusion profiles are produced such that $N_i^- = 0$ for all these elements. However, Fe diffusion profile has $N_i^- \neq 0$ or $N_i^+ \neq 0$. Therefore, as small error in detecting the location of the marker plane can give much higher difference in estimated diffusion coefficients (by calculating the intrinsic flux of Fe following Eq. 6), which can be seen in Table 4b. Still the diffusion couple produced in this study is very good, which can be understood by comparing the diffusion coefficient estimated at the Kirkendall marker plane and the self-diffusion coefficient of Fe (radiotracer diffusion coefficient of Fe in pure Fe) $D_{Fe}^* = 3.8 \times 10^{-15}\ m^2/s$ reported in literature. This is very close to the value estimated at the Kirkendall marker plane i.e. $2.6 \times 10^{-15}\ m^2/s$. This is a little lower than the self-diffusion coefficient and the diffusion coefficients of other elements are also found to be little lower than the impurity diffusion coefficients in Fe, which is discussed in the next section. Moreover, based on our experience in several analysis in different systems, we have noticed that estimating diffusion coefficients of the solvent following the Kirkaldy-Lane method can be even more uncertain since these are estimated from the interdiffusion coefficients, and the intrinsic and tracer diffusion coefficient of the solvent has little contribution to the interdiffusion process in a minor alloying element (solutes). The interdiffusion process is mainly controlled by the diffusion rates of the minor alloying elements, which can be understood from the Eq. 9. Rather success is found to be higher by estimating directly at the Kirkendall marker plane during analysis in different other systems, which will be published in future. Since the data estimated in this study could be compared with the impurity and self-diffusion coefficients available in literature and the impurity diffusion coefficients estimated in this study, we don't need to compare again by estimating the data following the Kirkaldy-Lane method, which is done in the Ni-Co-Fe system.

The diffusion couple prepared in Ni-Co-Fe-Cr-Al quinary system is shown in Fig. 7. The diffusion couple is prepared again such that $N_i^- = 0$ for Co, Fe, Cr and Al for minimum error in calculation of the intrinsic flux of these elements. The correlation between intrinsic flux and $n(n-1) = 20$ intrinsic diffusion coefficients in this quinary Ni-rich alloy system can be expressed as

$$V_m J_{Ni} = -D_{NiCo}^{Ni} \frac{\partial N_{Co}}{\partial x} - D_{NiFe}^{Ni} \frac{\partial N_{Fe}}{\partial x} - D_{NiCr}^{Ni} \frac{\partial N_{Cr}}{\partial x} - D_{NiAl}^{Ni} \frac{\partial N_{Al}}{\partial x} \tag{13a}$$

$$V_m J_{Co} = -D_{CoCo}^{Ni} \frac{\partial N_{Co}}{\partial x} - D_{CoFe}^{Ni} \frac{\partial N_{Fe}}{\partial x} - D_{CoCr}^{Ni} \frac{\partial N_{Cr}}{\partial x} - D_{CoAl}^{Ni} \frac{\partial N_{Al}}{\partial x} \tag{13b}$$

$$V_m J_{Fe} = -D_{FeCo}^{Ni} \frac{\partial N_{Co}}{\partial x} - D_{FeFe}^{Ni} \frac{\partial N_{Fe}}{\partial x} - D_{FeCr}^{Ni} \frac{\partial N_{Cr}}{\partial x} - D_{FeAl}^{Ni} \frac{\partial N_{Al}}{\partial x} \tag{13c}$$

$$V_m J_{Cr} = -D_{CrCo}^{Ni} \frac{\partial N_{Co}}{\partial x} - D_{CrFe}^{Ni} \frac{\partial N_{Fe}}{\partial x} - D_{CrCr}^{Ni} \frac{\partial N_{Cr}}{\partial x} - D_{CrAl}^{Ni} \frac{\partial N_{Al}}{\partial x} \tag{13d}$$

$$V_m J_{Al} = -D_{AlCo}^{Ni} \frac{\partial N_{Co}}{\partial x} - D_{AlFe}^{Ni} \frac{\partial N_{Fe}}{\partial x} - D_{AlCr}^{Ni} \frac{\partial N_{Cr}}{\partial x} - D_{AlAl}^{Ni} \frac{\partial N_{Al}}{\partial x} \tag{13e}$$

In this system, the composition of some of the alloying elements at the Kirkendall marker plane is significant to find significant contribution of the vacancy wind effect on certain cross intrinsic diffusion coefficients (which is explained in section 3.3). Therefore, the tracer diffusion coefficients are calculated considering the vacancy wind effect i.e. by replacing Eq. 7 in Eq. 13. It is evident from the diffusion profile of elements plotted in Fig. 5a that Co has a strong uphill diffusion profile in the alloy side of the diffusion couple. Aluminium diffusion profile is shown separately in Fig. 5b, which also has very strong uphill diffusion. These characteristics of diffusion profiles are discussed considering estimated intrinsic diffusion coefficients in Section 3.3. The location of the Kirkendall marker plane (K7) in this diffusion couple (DC7) is shown by a dotted line. The estimated tracer diffusion coefficients at this composition are listed in Table 5a. These are found to be similar to measured values reported earlier at another composition by intersecting two dissimilar pseudo-quaternary diffusion paths ($D^*_{Ni} = 5.1 \times 10^{-15}$, $D^*_{Co} = 5.4 \times 10^{-15}$, $D^*_{Fe} = 10.2 \times 10^{-15}$, $D^*_{Cr} = 14.4 \times 10^{-1}$ , $D^*_{Al} = 26.3 \times 10^{-15}\ m^2/s$) [22]. The values are similar since diffusion coefficients may not change significantly with the change in composition in a solid solution. The experimental effort and analysis at the Kirkendall marker plane of a single diffusion couple is much simpler and straightforward compared to previous analysis. Still, the need for more than one method to follow for estimation of the diffusion coefficients in a new multicomponent system cannot be denied at least one composition making sure the quality of the data generated. Every method has its advantages and disadvantages. In a multi principal alloy the estimation following constrained diffusion couple method [20, 22] is still more suitable when incremental diffusion couple need to be produced compared to the method proposed in this study more suitable in concentrated alloys such as Ni-, Co and Fe- based superalloys and steel used widely in various applications.

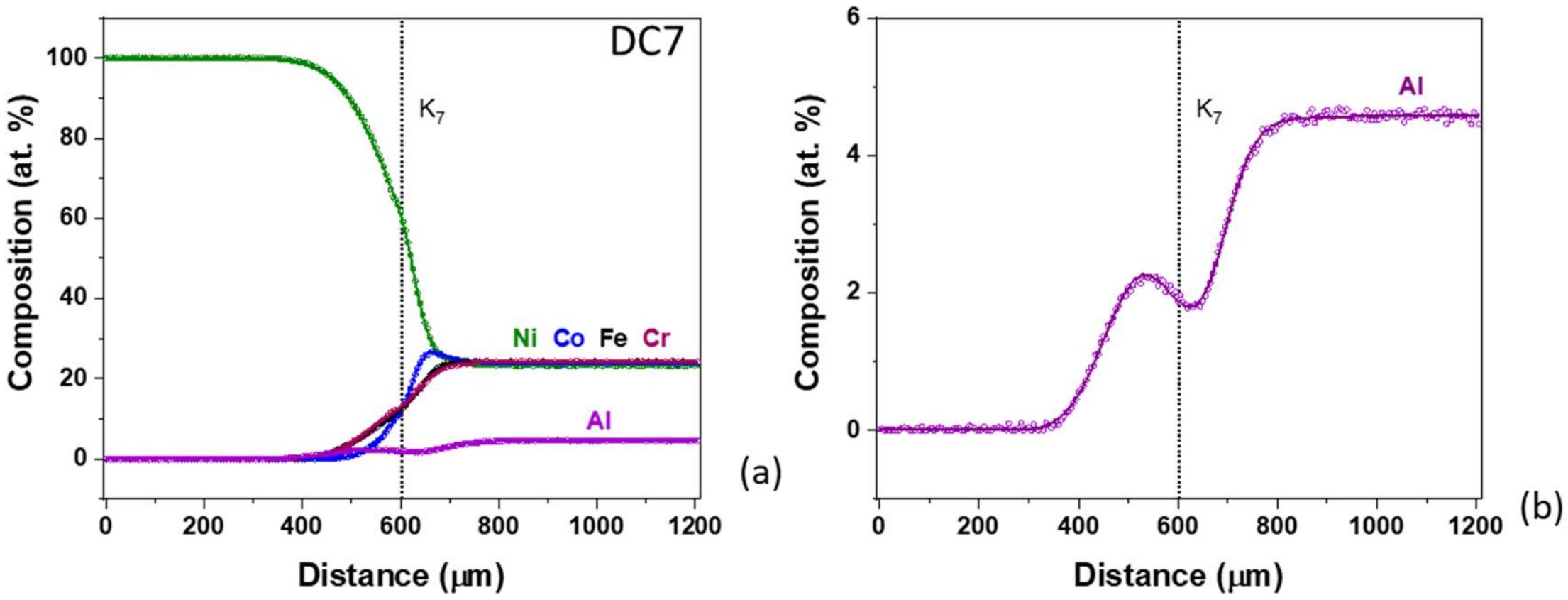

Fig. 5 (a) Diffusion profile, produce at 1200°C after annealing for 50 h. The Kirkendall marker plane location is indicated by $K_7$ (b) Al profile showing uphill near the location of Kirkendall marker plane.

As already discussed, we have again produced the diffusion couples such that $N_i^- = 0$ for Co, Fe, Cr, Al. However, the same could not be done for Ni. We have shown the difference in data estimated in Table 5b if the location of the Kirkendall marker plane is detected by uncertainty in analysis at $\pm 1\ \mu m$ and $\pm 2\ \mu m$ from the location detected. It can be seen that the difference in estimated data for Co, Fe, Cr and Al would make a very little difference. However, since $N_i^- \neq 0$ or $N_i^+ \neq 0$ for Ni, it produces a little higher difference in estimated data. Still the estimated data in this Ni-rich alloy is not affected as high as Fe in Fe-rich FeNiCoCr system (see Table 4b) because of higher composition range. The uncertainty of locating the Kirkendall marker plane and, therefore, the difference in data is also directly related to the length of the interdiffusion zone. Therefore, higher composition range of the diffusion couple and reasonably high diffusion annealing are the key for producing reliable diffusion coefficients following the incremental diffusion couples. These aspects are similar to the problems faced in binary diffusion couples as well [51-52].

Table 5 (a) Tracer diffusion coefficients estimated in diffusion couple D7 (Ni-rich, NiCoFeCrAl quinary alloy) at 1200°C (b) Error in calculation if the location of the Kirkendall marker plane (K7) is detected at $\pm 1$ or $\pm 2\ \mu m$ from the position detected in this study.

| Marker Plane Location Composition (at. %) | Tracer Diffusion Coefficients ($\times 10^{-15}\ m^2/s$) | | | | |
|---|---|---|---|---|---|
| | $D^*_{Ni}$ | $D^*_{Co}$ | $D^*_{Fe}$ | $D^*_{Cr}$ | $D^*_{Al}$ |
| DC 7 ($K_7$) (Fe = 12.5 Ni =59.9 Co = 12.5 Cr = 13.2 Al = 1.9 ) | 5.5±1 | 7.3±1.4 | 16.6±2.4 | 16.4±2.3 | 28.1±4.5 |

(a)

| Distance from K7 | Tracer Diffusion Coefficient (x $10^{-15}$ $m^2/s$) | | | | | % of error | | | | |
|---|---|---|---|---|---|---|---|---|---|---|
| | $D^*_{Ni}$ | $D^*_{Co}$ | $D^*_{Fe}$ | $D^*_{Cr}$ | $D^*_{Al}$ | $D^*_{Ni}$ | $D^*_{Co}$ | $D^*_{Fe}$ | $D^*_{Cr}$ | $D^*_{Al}$ |
| -2 µm | 4.8 | 7.1 | 16.2 | 16.2 | 27.4 | -11.8 | -3.3 | -2.1 | -1.6 | -2.5 |
| -1 µm | 5.1 | 7.2 | 16.4 | 16.3 | 27.9 | -5.9 | -1.8 | -0.8 | -0.5 | -0.6 |
| K7 position | 5.5 | 7.3 | 16.6 | 16.4 | 28.1 | - | - | - | - | - |
| + 1 µm | 5.8 | 7.4 | 16.8 | 16.6 | 28.6 | 6.3 | 1.3 | 1.5 | 0.8 | 2.0 |
| + 2 µm | 6.1 | 7.5 | 16.9 | 16.7 | 28.8 | 11.8 | 2.9 | 2.3 | 1.6 | 2.6 |

(b)

### *3.2 Estimation of impurity diffusion coefficients of elements following the Hall method*

Even the impurity diffusion coefficients of elements in pure component estimated can be very useful for the material systems with low concentration of alloying elements. Hall proposed a method of estimating the impurity diffusion coefficient of elements from the interdiffusion profile developed by the diffusion couple method [24,25]. He proposed an analytical expression, which can be used at the very low concentrations by probability plot of composition profile measured. It was proposed to estimate the diffusion coefficient in a binary system at the extreme composition range in which estimation of diffusion coefficients following the Matano-Boltzmann [58,59] or Sauer-Freise [43] method is difficult. This can also be extended to the ternary or multicomponent systems [60]. The proposed relations in Hall method are [25]

$$\frac{N_i-N_i^-}{N_i^+-N_i^-}=\frac{N_i}{N_i^+}=\frac{1}{2}(1+erfU) \qquad (14a)$$

Since $N_i^- = 0$ for the elements for which the impurity diffusion coefficients are estimated.

$$U=h\lambda+k \qquad (14b)$$

where, $\lambda=\frac{x-x_o}{t^{1/2}}$ is the Boltzmann parameter [59], $x_o$ is the initial contact plane of the diffusion couple or the Matano plane [58]. In this, composition vs. distance plot is converted to U vs. $\lambda$ linear plot to determine the values of $h$ (slope) and $k$ (intercept). Following, the impurity diffusion coefficient of the element is calculated from [25]

$$D=\frac{1}{4h^2}\left[1+k\sqrt{\pi}\exp{(U)^2}(1+erfU)\right] \qquad (14c)$$

For example, the U vs. $\lambda$ plots for estimation of impurity diffusion coefficients of Ni and Co in Fe from DC 1, diffusion coefficients of Co and Fe in Ni from DC 2, and impurity diffusion coefficients of Ni an Fe in Co from DC 3 are shown in Fig. 6. Similar analysis was conducted in quaternary Fe-FeNiCoCr and quinary Ni-NiCoFeCrAl systems as well (plots are given in the supplementary file, S4 and S5). The estimated data are listed in Table 6. The impurity diffusion coefficients of elements in Fe (Table 6b) from ternary and quaternary systems are found to be very close. Similarly, the impurity diffusion coefficient of elements in Ni (Table 6c) from ternary and quinary system diffusion couples are also found to be very close. The comparison of the diffusion coefficients estimated with the data available in literature, which were measured following different methods (radiotracer and by extending the binary interdiffusion coefficients to pure elements) is shown in Fig. 7. Therefore, as demonstrated, we can estimate all types of diffusion coefficients of all the elements at the Kirkendall marker plane and also the impurity diffusion coefficients of elements from a single diffusion profile in a multicomponent system.

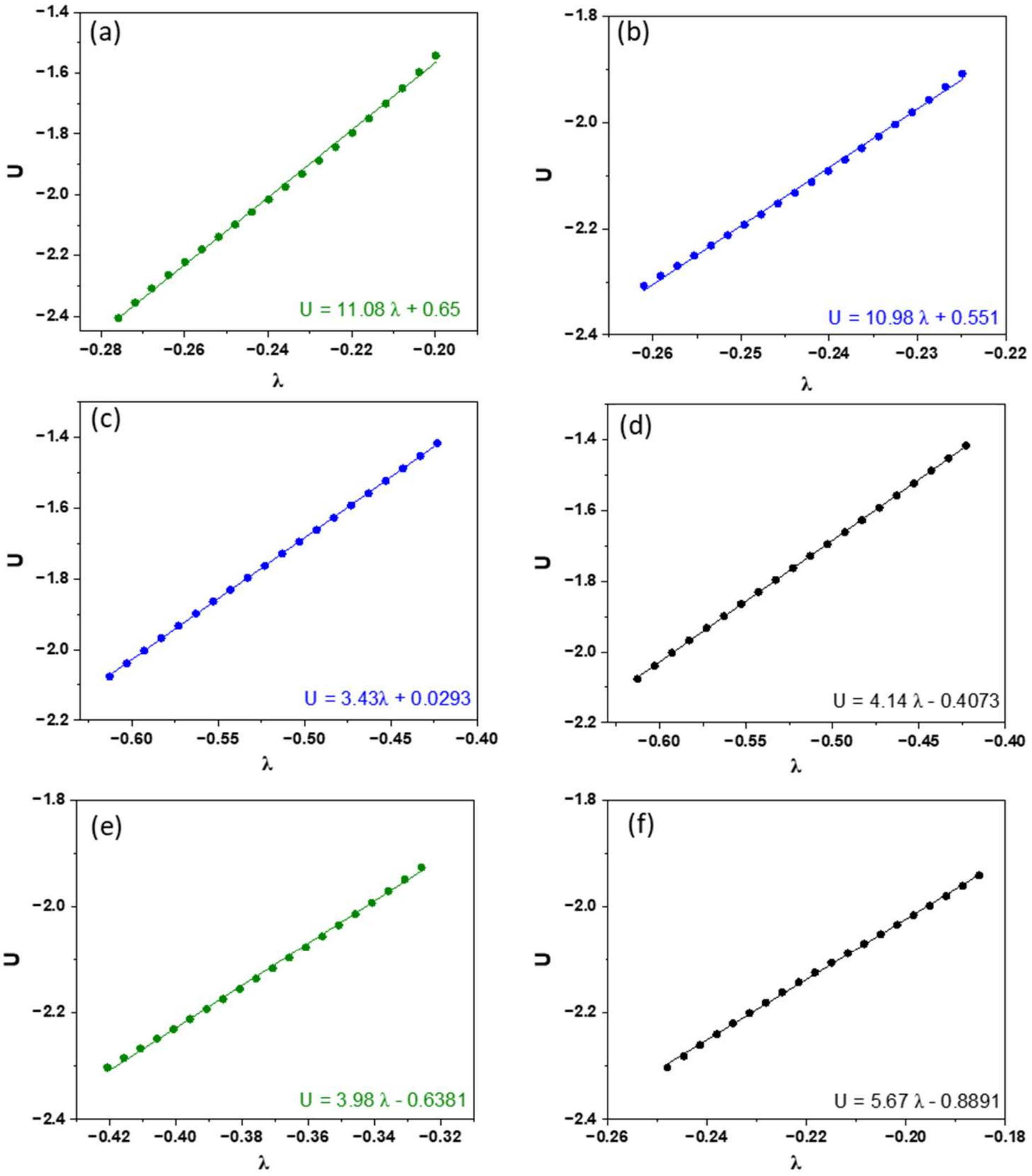

Fig.6 Estimation of impurity diffusion coefficients of (a) Ni , (b) Co in Fe from DC1, (c) Co (d) Fe in Ni from DC2 and, (e) Ni and (f) Fe in Co from DC3 at 1200°C.

Table 6: Impurity measurement using hall method in (a) Co, (b) Fe and (c) Ni

| | Ternary Co-NiFe (DC3) ($\times 10^{-15} m^2/s$) |
|---|---|
| $D^{imp}_{\mathrm{Ni\,(Co)}}$ | 5±1.5 |
| $D^{imp}_{\mathrm{Fe\,(Co)}}$ | 11.8±2.2 |

(a)

| | Fe-NiCo (DC1) ($\times 10^{-15} m^2/s$) | Fe-FeNiCoCr (DC6) ($\times 10^{-15} m^2/s$) |
|---|---|---|
| $D^{imp}_{\mathrm{Ni\,(Fe)}}$ | 2.6±0.55 | 2.5±0.5 |
| $D^{imp}_{\mathrm{Co\,(Fe)}}$ | 2.5±0.5 | 2.9±0.6 |
| $D^{imp}_{Cr(\mathrm{Fe})}$ | | 5.5±1 |

(b)

| | Ni-CoFe (DC2 ($\times 10^{-15} m^2/s$) | Ni-NiCoFeCrAl (DC7) ($\times 10^{-15} m^2/s$) |
|---|---|---|
| $D^{imp}_{Co\,(Ni)}$ | 11.9±2 | 11.4±2 |
| $D^{imp}_{Fe\,(Ni)}$ | 21.5±3.5 | 25.5±4 |
| $D^{imp}_{Cr(Ni)}$ | | 23.3±4.2 |
| $D^{imp}_{Al\,(Ni)}$ | | 53.1±9 |

(c)

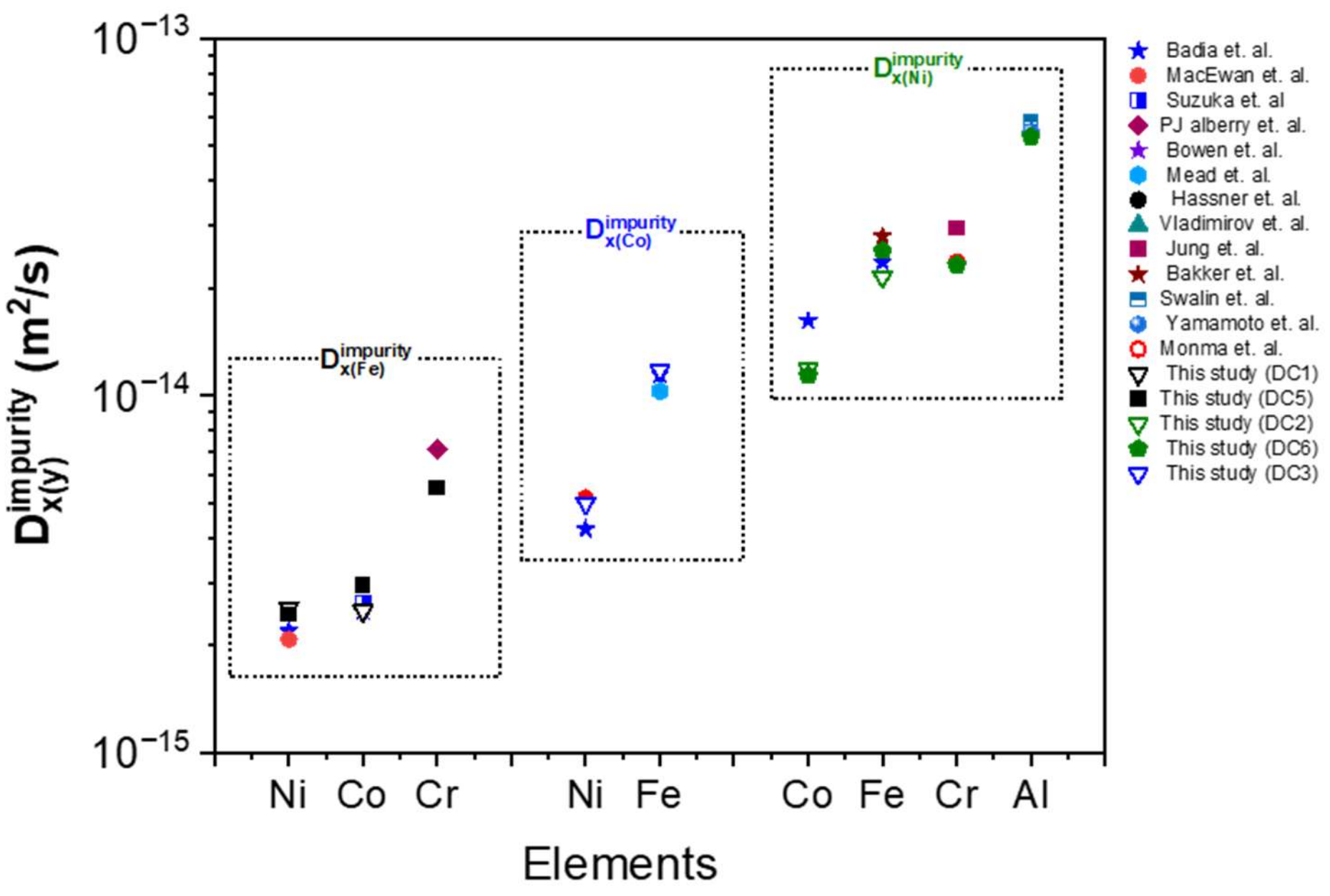

Fig. 7. Comparison of Impurity diffusion coefficient estimated using Hall method with the available literature data.

*3.3 Significance of estimating the intrinsic and interdiffusion coefficients*

As already mentioned in the introduction, most of the diffusion studies conducted in the ternary systems for estimation of the interdiffusion coefficients. Even the body diagonal method proposed for estimation diffusion coefficients in multicomponent systems was also for estimation of the interdiffusion coefficients. Kirkaldy and Lane [10] proposed the estimation tracer and then intrinsic diffusion coefficients from the estimated interdiffusion coefficients at the intersecting composition of two diffusion paths in a ternary system. This was proposed a long ago without any use of this method experimentally most probably because of unavailability of thermodynamic data available in ternary systems. The group of van Loo [61] first used this method much after in Cu-Ni-Fe system successfully for the estimation of the tracer diffusion coefficients and then proposed the correlations between tracer and intrinsic diffusion coefficients with thermodynamic factor instead of chemical potential gradient, which was proposed by Kirkaldy and Lane. Recently, we have shown the estimation of tracer, intrinsic, and interdiffusion coefficients by introducing the new concepts of intersecting similar [13,62] or dissimilar constrained diffusion paths [20-22] in multicomponent systems from two diffusion paths. The change in mindset to estimate the tracer diffusion coefficients directly from the interdiffusion flux instead of the interdiffusion coefficients first (including the approach proposed by Kirkaldy and Lane [10]) reduced to the need for only two intersecting or closely passed diffusion paths instead of (n-1) diffusion paths, which has an immense benefit especially in the multicomponent system.

Table 7 Calculated values of intrinsic , interdiffusion coefficients and vacancy wind effects at 1200°C at the Kirkendall marker plane composition of DC1 (K1) in Ni-Co-Fe ternary system

| $\boldsymbol{\phi}_{ij}^{n}$ | | $D_{ij}$ ($\times$ 10$^{-15}$ m$^2$/s ) | | | $1+W_{ij}$ | $\widetilde{D}_{ij}$ ($\times$ 10$^{-15}$ m$^2$/s ) | | | $1+\widetilde{W}_{ij}$ |
|---|---|---|---|---|---|---|---|---|---|
| | | | Without VWE | With VWE | | | Without VWE | With VWE | |
| $\boldsymbol{\phi}_{FeNi}^{Fe}$ | -0.5455 | $D_{FeNi}^{Fe}$ | -14.6 | -16.4 | 1.12 | | | | |
| $\boldsymbol{\phi}_{FeCo}^{Fe}$ | -0.5680 | $D_{FeCo}^{Fe}$ | -15.7 | -17.7 | 1.13 | | | | |
| $\boldsymbol{\phi}_{NiNi}^{Fe}$ | 1.0206 | $D_{NiNi}^{Fe}$ | 5.0 | 4.7 | 0.94 | $\widetilde{D}_{NiNi}^{Fe}$ | 7.1 | 7.3 | 1.03 |
| $\boldsymbol{\phi}_{CoNi}^{Fe}$ | 0.1889 | $D_{CoNi}^{Fe}$ | 0.9 | 0.5 | 0.56 | $\widetilde{D}_{CoNi}^{Fe}$ | 2.9 | 3.1 | 1.07 |
| $\boldsymbol{\phi}_{NiCo}^{Fe}$ | 0.1426 | $D_{NiCo}^{Fe}$ | 0.7 | 0.4 | 0.57 | $\widetilde{D}_{NiCo}^{Fe}$ | 3.0 | 3.3 | 1.10 |
| $\boldsymbol{\phi}_{CoCo}^{Fe}$ | 1.1569 | $D_{CoCo}^{Fe}$ | 5.4 | 5.1 | 0.94 | $\widetilde{D}_{CoCo}^{Fe}$ | 7.6 | 7.9 | 1.04 |

In this study, we have introduced the direct estimation of tracer diffusion coefficients in a multicomponent system, which needs only single diffusion couple. The estimation of the tracer diffusion coefficients is already introduced in section 3.1. Following, we can calculate the intrinsic following Eq. 7 and the interdiffusion coefficients following Eq. 9. The estimation of such data at the Kirkendall marker plane K1 in the Ni-Co-Fe are shown in Table 7. The estimated data at other marker planes are given in the supplementary file S6. Based on the analysis in the binary system, we know that the intrinsic diffusion coefficients of the faster element increase, and intrinsic diffusion coefficients of the slower element decreases because of the vacancy wind effect [3]. It can be understood from $(1+W_{ij})>1$ for intrinsic diffusion coefficients of Fe ($D_{Fej}^{Fe}$) but $\left(1+W_{ij}\right)<1$ for intrinsic diffusion coefficients of Ni ($D_{Nij}^{Fe}$) and Co ($D_{Coj}^{Fe}$) since we have $D_{Fe}^{*}>D_{Co}^{*}\approx D_{Ni}^{*}$. The cross intrinsic diffusion coefficients of Ni ($D_{NiCo}^{Fe}$) and Co ($D_{CoNi}^{Fe}$) are affected most although these are much smaller than the main intrinsic diffusion coefficients ($D_{NiNi}^{Fe}$) and ($D_{CoCo}^{Fe}$) such that $\frac{D_{NiCo}^{Fe}}{D_{NiNi}^{Fe}}=0.11$ and $\frac{D_{CoNi}^{Fe}}{D_{CoCo}^{Fe}}=0.08$. Moreover, the cross intrinsic diffusion coefficients of Fe have much higher value and with opposite sign compared to the cross intrinsic diffusion coefficients of Ni and Co such that $\frac{D_{FeCo}^{Fe}}{D_{NiCo}^{Fe}}=-44.25$ and $\frac{D_{FeNi}^{Fe}}{D_{CoNi}^{Fe}}=-32.5$. The main intrinsic diffusion coefficients in this type of solid solution are generally positive. In reference to Eq. 5, since the cross-diffusion coefficients of both Ni and Co are positive (positive diffusional interaction), the flux increases when these elements diffuse with same sign of composition gradients (same direction). Flux decreases if these diffuse with opposite sign of composition gradient (opposite direction). However, this change depending on the diffusion direction is expected to be marginal since the cross-diffusion coefficient values of these two elements are much lower compared to the main intrinsic diffusion coefficients. On the other hand, the cross intrinsic diffusion coefficients of Fe with Ni and Co have negative sign (negative diffusional interaction). It means that the flux of Fe will enhance when Fe diffuses in the direction with opposite sign of composition gradient compared to Ni and Co (opposite direction). However, the flux of Fe will reduce if Fe diffuses with same sign of composition gradient of Ni or Co (same direction). The uphill diffusion profiles of Co and Ni in DC 2 and DC 3 are evident since the diffusion couples are produced by forcing Ni and Co in these two diffusion couples to diffusion in the same direction from FeCo (DC 2) and NiCo (DC3) couples.

Let us now compare the interdiffusion coefficients to examine what we understand from these diffusion parameters compared to the intrinsic diffusion coefficients. From Eq. 9, it must be clear that every interdiffusion coefficient in a ternary system is a kind of average of a particular set of three intrinsic diffusion coefficients. The vacancy wind effect on all the interdiffusion coefficients is found to smaller than the similar main and cross intrinsic diffusion coefficients. However, these vacancy wind effects (of interdiffusion coefficients) are kind of vague terms since these are kind of average in a very complex manner of a set of three intrinsic diffusion coefficients. Interestingly, it can be noticed that the ratio of cross to main interdiffusion coefficients are $\frac{\tilde{D}_{NiCo}^{Fe}}{\tilde{D}_{NiNi}^{Fe}}=0.42$ and $\frac{\tilde{D}_{CoNi}^{Fe}}{\tilde{D}_{CoCo}^{Fe}}=0.42$. It's already discussed earlier that the ratio of similar intrinsic diffusion coefficients is much lower than these values (0.11 and 0.08, respectively). The cross interdiffusion coefficients are related to a set of three intrinsic diffusion coefficients in this ternary system following Eq. 9 as

$$\tilde{D}_{NiCo}^{Fe}=(1-N_{Ni})D_{NiCo}^{Fe}-N_{Ni}(D_{CoCo}^{Fe}+D_{FeCo}^{Fe}) \quad (15a)$$

$$\tilde{D}_{CoNi}^{Fe}=(1-N_{Co})D_{CoNi}^{Fe}-N_{Co}(D_{NiNi}^{Fe}+D_{FeNi}^{Fe}) \quad (15b)$$

Since the cross intrinsic diffusion coefficients of Fe ($D_{FeNi}^{Fe}$ and $D_{FeCo}^{Fe}$) have much higher values with negative sign, cross interdiffusion coefficients $\tilde{D}_{NiCo}^{Fe}$ and $\tilde{D}_{CoNi}^{Fe}$ have much higher value compared to the cross intrinsic diffusion coefficients $D_{NiCo}^{Fe}$ and $D_{CoNi}^{Fe}$. Therefore, if we estimate only the interdiffusion coefficients directly at the intersecting composition without knowing the intrinsic diffusion coefficients, we may have misleading information to say wrongly that the Ni and Co have strong diffusional interactions instead of weak diffusional interactions.

The discussion above indicates that intrinsic diffusion coefficients are very important to estimate for knowing the actual diffusional interactions between the elements. The composition range considered in the analysis have significant concentration of all the element. The composition at which the diffusion coefficients are estimated is important for this comparison of intrinsic and interdiffusion coefficients (see Eq. 9). Now let us examine the estimated diffusion coefficients in a Fe-rich quaternary alloy at a composition with minor concentration of Ni, Co and Cr. The estimated tracer diffusion coefficients at the Kirkendall marker plane composition are listed in Table 4a. The intrinsic diffusion coefficients are calculated considering Eq. 7. These are calculated considering and neglecting ($W_{ij} = 0$) the vacancy wind effect. Following, the interdiffusion coefficients are calculated considering Eq. 9, as listed in Table 8. It can be seen that in this Fe concentrated alloy, the vacancy wind effect has little influence on most of the intrinsic diffusion coefficients except certain cross intrinsic diffusion coefficients with negligible values. There is no influence of the vacancy wind effect again on most of the interdiffusion coefficients. Therefore, analysis can be conducted without considering the vacancy wind effect. We found this is true for other (Ni- and Fe-rich) concentrated alloys as well (based on yet to be published research). Another important fact to be noticed here that the main and cross interdiffusion and intrinsic diffusion coefficients have same values up to one decimal point since the composition of the alloying elements (solutes), $N_i$ are small (see Eq. 9). This further means that the tracer diffusion coefficients can be calculated directly from the main interdiffusion coefficients since $\widetilde{D}_{ii}^{k} \approx D_{ii}^{k} \approx D_i^{*}\emptyset_{ii}^{k}$. One can estimate tracer diffusion coefficients of all the elements by calculating the interdiffusion coefficients considering different element as the dependent variable. Therefore, estimating the interdiffusion coefficients directly is enough to understand the diffusional interactions in such systems.

Table 8 Calculated values of intrinsic , interdiffusion coefficients and vacancy wind effects at 1200°C at the Kirkendall marker plane composition of DC6 (K6) in Fe-rich Fe-Ni-Co-Cr quaternary system

| $\boldsymbol{\phi}_{ij}^{n}$ | | $D_{ij}$ ($\times 10^{-15}$ m $^2$/s) | | | $1 + W_{ij}$ | $\widetilde{D}_{ij}$ ($\times 10^{-15}$ m $^2$/s) | | | $1 + \widetilde{W}_{ij}$ |
|---|---|---|---|---|---|---|---|---|---|
| | | | Without VWE | With VWE | | | Without VWE | With VWE | |
| $\boldsymbol{\phi}_{FeNi}^{Fe}$ | -0.0477 | $D_{FeNi}^{Fe}$ | -2.6 | -2.7 | 1.04 | | | | |
| $\boldsymbol{\phi}_{FeCo}^{Fe}$ | -0.0531 | $D_{FeCo}^{Fe}$ | -2.7 | -2.9 | 1.07 | | | | |
| $\boldsymbol{\phi}_{FeCr}^{Fe}$ | -0.0399 | $D_{FeCr}^{Fe}$ | -2.6 | -1.9 | 0.73 | | | | |
| $\boldsymbol{\phi}_{NiNi}^{Fe}$ | 0.9856 | $D_{NiNi}^{Fe}$ | 1.9 | 1.9 | 1.00 | $\widetilde{D}_{NiNi}^{Fe}$ | 1.9 | 1.9 | 1.00 |
| $\boldsymbol{\phi}_{CoNi}^{Fe}$ | 0.0276 | $D_{CoNi}^{Fe}$ | 0.1 | 0.1 | 1.00 | $\widetilde{D}_{CoNi}^{Fe}$ | 0.1 | 0.1 | 1.00 |
| $\boldsymbol{\phi}_{CrNi}^{Fe}$ | -0.0219 | $D_{CrNi}^{Fe}$ | -0.1 | -0.1 | 1.00 | $\widetilde{D}_{CrNi}^{Fe}$ | -0.1 | -0.1 | 1.00 |
| $\boldsymbol{\phi}_{NiCo}^{Fe}$ | 0.0267 | $D_{NiCo}^{Fe}$ | 0.05 | 0.04 | 0.80 | $\widetilde{D}_{NiCo}^{Fe}$ | 0.1 | 0.1 | 1.00 |
| $\boldsymbol{\phi}_{CoCo}^{Fe}$ | 1.0369 | $D_{CoCo}^{Fe}$ | 2.1 | 2.1 | 1.00 | $\widetilde{D}_{CoCo}^{Fe}$ | 2.1 | 2.1 | 1.00 |
| $\boldsymbol{\phi}_{CrCo}^{Fe}$ | -0.0188 | $D_{CrCo}^{Fe}$ | -0.1 | -0.1 | 1.00 | $\widetilde{D}_{CrCo}^{Fe}$ | -0.1 | -0.1 | 1.00 |
| $\boldsymbol{\phi}_{NiCr}^{Fe}$ | -0.0179 | $D_{NiCr}^{Fe}$ | -0.04 | -0.02 | 0.50 | $\widetilde{D}_{NiCr}^{Fe}$ | -0.2 | -0.2 | 1.00 |
| $\boldsymbol{\phi}_{CoCr}^{Fe}$ | -0.0126 | $D_{CoCr}^{Fe}$ | -0.03 | -0.01 | 0.33 | $\widetilde{D}_{CoCr}^{Fe}$ | -0.1 | -0.2 | 2.00 |
| $\boldsymbol{\phi}_{CrCr}^{Fe}$ | 1.0245 | $D_{CrCr}^{Fe}$ | 5.2 | 5.3 | 1.02 | $\widetilde{D}_{CrCr}^{Fe}$ | 5.2 | 5.2 | 1.00 |

Now let us discuss the intrinsic and interdiffusion coefficients estimated at the Kirkendall marker plane position in the NiCoFeCrAl system in which all the alloying elements (except Al) have composition greater than 10 at.%, as mention in Table 5a along with the estimated tracer diffusion coefficients. Table 9a lists the intrinsic diffusion coefficients considering and neglecting the vacancy wind effects for Ni as the dependent variable. It can be seen that certain cross intrinsic diffusion coefficients (such as $D_{NiFe}^{Ni}$, $D_{NiCr}^{Ni}$, $D_{NiAl}^{Ni}$, $D_{FeAl}^{Ni}$, $D_{CrAl}^{Ni}$) with reasonable values (not negligible compared to the main intrinsic diffusion coefficients) are affected by the vacancy wind effect by $\geq 20\%$ although the vacancy wind effect are smaller for most of similar interdiffusion coefficients. This is opposite, for example, for $D_{FeAl}^{Ni}$ and $\widetilde{D}_{FeAl}^{Ni}$. Therefore, interdiffusion coefficients do not reflect on the actual vacancy wind effect since these are kind of average of a set of intrinsic diffusion coefficients (five in a quinary system). Moreover, sometimes, the intrinsic and interdiffusion coefficients have very different values ( for example, $D_{CoFe}^{Ni}$ and $\widetilde{D}_{CoFe}^{Ni}$, $D_{FeCr}^{Ni}$ and $\widetilde{D}_{FeCr}^{Ni}$, $D_{FeAl}^{Ni}$ and $\widetilde{D}_{FeAl}^{Ni}$). These are similar for $D_{Alx}^{Ni}$ a $\widetilde{D}_{Alx}^{Ni}$ since the composition of Al is relatively small (See Eq. 9). Therefore, the interdiffusion coefficients may not reflect on the diffusional interactions correctly if these are only estimated, let say by passing (n-1) diffusion paths, and the intrinsic diffusion coefficients are not estimated.

Another important point should be noted here. Sometimes, the main interdiffusion coefficients are only estimated to make a comment on the relative mobilities of the elements assuming these will project a similar trend compared to the tracer diffusion coefficients. Let us firsts refer to the interdiffusion coefficients estimated in the Ni-Co-Fe system. The data estimated considering Fe are listed in Table 7 and the data estimated considering Ni and Co are also listed in the supplementary file (S7). It can be seen that $(\widetilde{D}_{FeFe}^{Ni}\ or\ \widetilde{D}_{FeFe}^{Co}) > (\widetilde{D}_{CoCo}^{Ni}\ or\ \widetilde{D}_{CoCo}^{Co}) \approx (\widetilde{D}_{NiNi}^{Co}\ or\ \widetilde{D}_{NiNi}^{Fe})$. Therefore, these diffusion parameters correctly indicates on the relative mobilities of the since we have $D_{Fe}^{*} > D_{Co}^{*} \approx D_{Ni}^{*}$ although the values of the tracer and intrinsic diffusion coefficients are different. However, the interdiffusion coefficients considering different elements as the dependent variable since $\widetilde{D}_{ij}^{n} = \widetilde{D}_{ij} - \widetilde{D}_{in}$ are always relative values. This may cause a greater confusion in certain systems. Let us consider the interdiffusion coefficients estimated in the Ni-Co-Fe-Cr-Al system considering Ni and Al as the dependent variable as listed in Table 9a and 9b. It can be seen that, the thermodynamic factors, intrinsic and interdiffusion coefficients are different when we consider Ni or Al as the dependent variables. Moreover, we have $\widetilde{D}_{CoCo}^{Ni} < \widetilde{D}_{FeFe}^{Ni} \leq \widetilde{D}_{CrCr}^{Ni} < \widetilde{D}_{AlAl}^{Ni}$ considering Ni as the dependent variable; however, $\widetilde{D}_{NiNi}^{Al} > \widetilde{D}_{CoCo}^{Al} < \widetilde{D}_{FeFe}^{Al} > \widetilde{D}_{CrCr}^{Al}$ considering Al as the dependent variable. It is true that the interdiffusion flux (diffusion profiles) can be related correctly for interdiffusion based analysis with interdiffusion coefficients considering any one of the elements as the dependent variables; however, these terms are vague and confusing when we try to get an indication on the diffusional interactions and relative mobilities of the elements. It is very important to estimate the tracer and intrinsic diffusion coefficients also need to check data considering different elements as the dependent values very carefully before making any statement on the atomic mechanism of diffusion especially multi-principal element alloys or alloys with relatively high concentration of alloying elements (solutes).

Table 9 Calculated values of intrinsic , interdiffusion coefficients and vacancy wind effects at 1200°C at the Kirkendall marker plane composition of DC7 (K7) in Ni-rich Ni-Co-Fe-Cr-Al quinary system (a) considering Ni as the dependent variable and (b) considering Al as the dependent variable.

| $\phi_{ij}^{n}$ | | | $D_{ij}$ ($\times 10^{-15}$ m$^2$/s) | | $1+W_{ij}$ | | $\tilde{D}_{ij}$ ($\times 10^{-15}$ m$^2$/s) | | $1+\tilde{W}_{ij}$ |
|---|---|---|---|---|---|---|---|---|---|
| | | | Without VWE | With VWE | | | Without VWE | With VWE | |
| $\phi_{NiCo}^{Ni}$ | -0.2169 | $D_{NiCo}^{Ni}$ | -5.7 | -5.4 | 0.95 | | | | |
| $\phi_{NiFe}^{Ni}$ | -0.3069 | $D_{NiFe}^{Ni}$ | -8.0 | -6.2 | 0.78 | | | | |
| $\phi_{NiCr}^{Ni}$ | -0.3582 | $D_{NiCr}^{Ni}$ | -8.9 | -6.8 | 0.76 | | | | |
| $\phi_{NiAl}^{Ni}$ | -0.0854 | $D_{NiAl}^{Ni}$ | -15.0 | -10.4 | 0.69 | | | | |
| $\phi_{CoCo}^{Ni}$ | 0.9758 | $D_{CoCo}^{Ni}$ | 7.1 | 7.2 | 1.01 | $\tilde{D}_{CoCo}^{Ni}$ | 6.8 | 6.7 | 0.99 |
| $\phi_{FeCo}^{Ni}$ | 0.0245 | $D_{FeCo}^{Ni}$ | 0.4 | 0.6 | 1.50 | $\tilde{D}_{FeCo}^{Ni}$ | 0.0 | 0.1 | -- |
| $\phi_{CrCo}^{Ni}$ | -0.0198 | $D_{CrCo}^{Ni}$ | -0.3 | -0.1 | 0.33 | $\tilde{D}_{CrCo}^{Ni}$ | -0.7 | -0.7 | 1.00 |
| $\phi_{AlCo}^{Ni}$ | 0.3708 | $D_{AlCo}^{Ni}$ | 1.6 | 1.6 | 1.00 | $\tilde{D}_{AlCo}^{Ni}$ | 1.5 | 1.5 | 1.00 |
| $\phi_{CoFe}^{Ni}$ | -0.0592 | $D_{CoFe}^{Ni}$ | -0.4 | 0.1 | -0.25 | $\tilde{D}_{CoFe}^{Ni}$ | -2.7 | -2.8 | 1.04 |
| $\phi_{FeFe}^{Ni}$ | 1.1770 | $D_{FeFe}^{Ni}$ | 19.5 | 20.7 | 1.06 | $\tilde{D}_{FeFe}^{Ni}$ | 17.3 | 17.8 | 1.03 |
| $\phi_{CrFe}^{Ni}$ | 0.2263 | $D_{CrFe}^{Ni}$ | 3.9 | 5.1 | 1.31 | $\tilde{D}_{CrFe}^{Ni}$ | 1.5 | 2.1 | 1.40 |
| $\phi_{AlFe}^{Ni}$ | 0.7521 | $D_{AlFe}^{Ni}$ | 3.1 | 3.4 | 1.10 | $\tilde{D}_{AlFe}^{Ni}$ | 2.8 | 3.0 | 1.07 |
| $\phi_{CoCr}^{Ni}$ | -0.1501 | $D_{CoCr}^{Ni}$ | -1.0 | -0.4 | 0.40 | $\tilde{D}_{CoCr}^{Ni}$ | -3.7 | -3.8 | 1.03 |
| $\phi_{FeCr}^{Ni}$ | 0.2069 | $D_{FeCr}^{Ni}$ | 3.3 | 4.6 | 1.39 | $\tilde{D}_{FeCr}^{Ni}$ | 0.6 | 1.3 | 2.17 |
| $\phi_{CrCr}^{Ni}$ | 1.4100 | $D_{CrCr}^{Ni}$ | 23.1 | 24.5 | 1.06 | $\tilde{D}_{CrCr}^{Ni}$ | 20.3 | 21.0 | 1.03 |
| $\phi_{AlCr}^{Ni}$ | 1.1773 | $D_{AlCr}^{Ni}$ | 4.7 | 5.0 | 1.06 | $\tilde{D}_{AlCr}^{Ni}$ | 4.3 | 4.5 | 1.05 |
| $\phi_{CoAl}^{Ni}$ | 0.0029 | $D_{CoAl}^{Ni}$ | 0.1 | 1.4 | 14.00 | $\tilde{D}_{CoAl}^{Ni}$ | -5.5 | -5.8 | 1.05 |
| $\phi_{FeAl}^{Ni}$ | 0.0716 | $D_{FeAl}^{Ni}$ | 8.0 | 10.9 | 1.36 | $\tilde{D}_{FeAl}^{Ni}$ | 2.3 | 3.7 | 1.61 |
| $\phi_{CrAl}^{Ni}$ | 0.1325 | $D_{CrAl}^{Ni}$ | 15.3 | 18.4 | 1.20 | $\tilde{D}_{CrAl}^{Ni}$ | 9.4 | 10.8 | 1.15 |
| $\phi_{AlAl}^{Ni}$ | 1.3052 | $D_{AlAl}^{Ni}$ | 36.7 | 37.4 | 1.02 | $\tilde{D}_{AlAl}^{Ni}$ | 35.8 | 36.3 | 1.01 |

(a)

| $\Phi_{ij}^{n}$ | | $D_{ij}$ ($\times$ 10$^{-15}$ m$^2$/s) | | | $1+W_{ij}$ | $\widetilde{D}_{ij}$ ($\times$ 10$^{-15}$ m$^2$/s) | | | $1+\widetilde{W}_{ij}$ |
|---|---|---|---|---|---|---|---|---|---|
| | | | Without VWE | With VWE | | | Without VWE | With VWE | |
| $\Phi_{AlNi}^{Al}$ | -42.2209 | $D_{AlNi}^{Al}$ | -37.4 | -38.2 | 1.02 | | | | |
| $\Phi_{AlCo}^{Al}$ | -8.4040 | $D_{AlCo}^{Al}$ | -35.8 | -36.5 | 1.02 | | | | |
| $\Phi_{AlFe}^{Al}$ | -8.1377 | $D_{AlFe}^{Al}$ | -34.4 | -34.9 | 1.01 | | | | |
| $\Phi_{AlCr}^{Al}$ | -8.0353 | $D_{AlCr}^{Al}$ | -32.4 | -32.8 | 1.01 | | | | |
| $\Phi_{NiNi}^{Al}$ | 2.7630 | $D_{NiNi}^{Al}$ | 15.4 | 10.7 | 0.69 | $\widetilde{D}_{NiNi}^{Al}$ | 42.9 | 46.0 | 1.07 |
| $\Phi_{CoNi}^{Al}$ | -0.1372 | $D_{CoNi}^{Al}$ | -0.2 | -1.5 | 7.50 | $\widetilde{D}_{CoNi}^{Al}$ | 5.5 | 5.8 | 1.05 |
| $\Phi_{FeNi}^{Al}$ | -2.0248 | $D_{FeNi}^{Al}$ | -7.3 | -10.4 | 1.42 | $\widetilde{D}_{FeNi}^{Al}$ | -1.6 | -3.0 | 1.88 |
| $\Phi_{CrNi}^{Al}$ | -4.5220 | $D_{CrNi}^{Al}$ | -16.4 | -19.5 | 1.19 | $\widetilde{D}_{CrNi}^{Al}$ | -10.3 | -11.7 | 1.14 |
| $\Phi_{NiCo}^{Al}$ | 0.3627 | $D_{NiCo}^{Al}$ | 9.7 | 5.3 | 0.55 | $\widetilde{D}_{NiCo}^{Al}$ | 35.4 | 38.2 | 1.08 |
| $\Phi_{CoCo}^{Al}$ | 0.9426 | $D_{CoCo}^{Al}$ | 7.0 | 5.8 | 0.83 | $\widetilde{D}_{CoCo}^{Al}$ | 12.3 | 12.6 | 1.02 |
| $\Phi_{FeCo}^{Al}$ | -0.4216 | $D_{FeCo}^{Al}$ | -7.3 | -10.2 | 1.40 | $\widetilde{D}_{FeCo}^{Al}$ | -1.9 | -3.3 | 1.74 |
| $\Phi_{CrCo}^{Al}$ | -0.9480 | $D_{CrCo}^{Al}$ | -16.5 | -19.4 | 1.18 | $\widetilde{D}_{CrCo}^{Al}$ | -10.9 | -12.1 | 1.11 |
| $\Phi_{NiFe}^{Al}$ | 0.2782 | $D_{NiFe}^{Al}$ | 7.4 | 4.6 | 0.62 | $\widetilde{D}_{NiFe}^{Al}$ | 23.8 | 25.6 | 1.08 |
| $\Phi_{CoFe}^{Al}$ | -0.1101 | $D_{CoFe}^{Al}$ | -0.8 | -1.6 | 2.00 | $\widetilde{D}_{CoFe}^{Al}$ | 2.6 | 2.8 | 1.08 |
| $\Phi_{FeFe}^{Al}$ | 0.7545 | $D_{FeFe}^{Al}$ | 13.1 | 11.2 | 0.85 | $\widetilde{D}_{FeFe}^{Al}$ | 16.5 | 15.6 | 0.95 |
| $\Phi_{CrFe}^{Al}$ | -0.7230 | $D_{CrFe}^{Al}$ | -12.5 | -14.3 | 1.14 | $\widetilde{D}_{CrFe}^{Al}$ | -8.9 | -9.7 | 1.09 |
| $\Phi_{NiCr}^{Al}$ | 0.2504 | $D_{NiCr}^{Al}$ | 6.4 | 3.8 | 0.59 | $\widetilde{D}_{NiCr}^{Al}$ | 21.2 | 22.8 | 1.08 |
| $\Phi_{CoCr}^{Al}$ | -0.1658 | $D_{CoCr}^{Al}$ | -1.2 | -1.9 | 1.58 | $\widetilde{D}_{CoCr}^{Al}$ | 1.9 | 2.1 | 1.11 |
| $\Phi_{FeCr}^{Al}$ | -0.2487 | $D_{FeCr}^{Al}$ | -4.1 | -5.7 | 1.39 | $\widetilde{D}_{FeCr}^{Al}$ | -1.0 | -1.8 | 1.80 |
| $\Phi_{CrCr}^{Al}$ | 0.3988 | $D_{CrCr}^{Al}$ | 6.6 | 4.9 | 0.74 | $\widetilde{D}_{CrCr}^{Al}$ | 9.8 | 9.1 | 0.93 |

(b)

## 4. Conclusion

It was a textbook knowledge until recently that estimating diffusion coefficients following the most practiced diffusion couple method is impossible in a system more than three elements. During the last one decade a few methods were proposed facilitating the estimation in a multicomponent system by intersecting or closely passing similar or dissimilar conventional [11,13] or constrained diffusion paths [19–23,62]. However, still, we need at least two diffusion paths to intersect in multicomponent space that needs very accurate control of diffusion couple end member compositions. By solving these problems, we have proposed another method for estimation of diffusion coefficients from a single diffusion profile. We have demonstrated this in ternary, quaternary, and quinary alloy systems. The conclusions made in this study are:

- We have proposed a method estimation of all types of diffusion coefficients from a single diffusion couple at the Kirkendall marker plane by changing the mindset of estimating the tracer diffusion coefficients first instead of the interdiffusion coefficients. This can be followed in a multicomponent system with any number of elements. Moreover, impurity diffusion coefficients of the elements can be estimated if a pure element is used as one of the end members of the diffusion couple.

- The interdiffusion coefficients are important for correlating with the diffusion paths or understanding the interdiffusion controlled processes. However, as demonstrated in this, these diffusion parameters are not enough or can be misleading in a composition range of a system with relatively high concentration of alloying elements explaining the importance of estimating tracer and intrinsic diffusion coefficients. Sometimes making an idea about the relative mobilities of the elements by comparing the main interdiffusion coefficients can be confusing, as shown in the Ni-Co-Fe-Cr-Al system.
- This design strategy of diffusion couples for estimation of diffusion coefficients with small error from a single diffusion couple is discussed in detail. This method can be very suitable for generating mobility database in Ni-, Co- or Fe-based alloys currently used in most of the structural and high temperature applications. However, this can be challenging in a multi-principal element alloy system when the diffusion couples cannot be prepared with one of the pure elements at one end or zero composition of an element in one of the end members. Incremental diffusion couple with non-zero composition at both ends of the diffusion couple may be tricky to estimate data with logical values avoiding relatively large error in calculation. In such a situation, it is advisable to practice the constrained diffusion couple methods by intersecting two diffusion paths for comparison of the data [20-22].
- On the other hand, based on our experience, it is advisable to practice two different methods at least certain at composition for generation of mobility database over a wide composition range. Different methods have different level and different sources of errors and comparable data from two different methods brings confidence in generation of mobility database. This is shown in the ternary system. In a system with minor alloying, the data estimated can be compared with the impurity diffusion coefficients, which is done in this study in the quaternary Fe-rich FeNiCoCr system. The data estimated in the NiCoFeCrAl system is compared with the data reported earlier which was measured by intersecting dissimilar constrained diffusion paths [22]. In general, diffusion coefficients may not change very significantly in a particular solid solution making it easier for comparison by estimating data at different compositions and/or by different methods. The reported data in this study and the previous study are indeed found to be comparable. Even the relative values of impurity diffusion coefficients in Ni found to follow a similar relative mobilities between the elements estimated at the Kirkendall marker plane.
- Based on the analysis in this study and reported earlier [22], we can state that the vacancy wind effect should not be neglected in concentrated or multi principal element alloys for accurate analysis.
- The mobility database in multicomponent systems generated purely experimentally is yet not available except at very few compositions, such as equiatomic Ni-Co-Fe-Cr, Ni-Co-Fe-Cr-Mn etc. The method proposed in this study along with the constrained diffusion couple method in combination to the numerical methods [63,64] can help generation of mobility database in various Ni-, Co- and Fe-based systems of immense practical importance to corelate with microstructural evolution and high temperature mechanical properties, which seemed impossible until recently.

**Acknowledgements:** We acknowledge the financial support from SERB, India (Project No. CRG/2021/001842).